\documentclass[preprint,showpacs,preprintnumbers,amsmath,amssymb]{revtex4}

\usepackage{graphicx}
\usepackage{dcolumn}
\usepackage{bm}

\begin{document}


\title{Disk Accretion Flow Driven by Large-Scale Magnetic Fields: Solutions
with Constant Specific Energy}

\author{Li-Xin Li\footnote{Chandra Fellow}}
  \affiliation{Harvard-Smithsonian Center for Astrophysics, Cambridge,
     MA 02138}

\date{\today}

\begin{abstract}
We study the dynamical
evolution of a stationary, axisymmetric, and perfectly conducting cold
accretion disk containing a large-scale magnetic field around a Kerr
black hole, trying to understand the relation between accretion and 
the transportation of angular momentum and energy. A one-dimensional 
radial momentum equation is derived near the equatorial plane, which 
has one intrinsic singularity at the fast critical point. We solve 
the radial momentum equation for solutions corresponding to an 
accretion flow that starts from a subsonic state at infinity, smoothly 
passes the fast critical point, then supersonically falls into the 
horizon of the black hole. The solutions always have the 
following features: 1)~The specific energy of fluid particles remains 
constant but the specific angular momentum is effectively removed by 
the magnetic field. 2)~At large radii, where the disk motion is 
dominantly rotational, the energy density of the magnetic field is 
equipartitioned with the rotational energy density of the disk. 
3)~Inside the fast critical point, where radial motion becomes 
important, the ratio of the electromagnetic energy density to 
the kinetic energy density drops quickly. The results indicate 
that: 1)~Disk accretion does not necessarily imply energy dissipation 
since magnetic fields do not have to transport or dissipate a lot of 
energy as they effectively transport angular momentum. 2)~When 
resistivity is small, the large-scale magnetic field is 
amplified by the shearing rotation of the disk until the magnetic energy 
density is equipartitioned with the rotational energy density, ending up
with a geometrically thick disk. This is in contrast with the evolution
of small-scale magnetic fields where if the resistivity is nonzero the 
magnetic energy density is likely to be equipartitioned with the kinetic 
energy density associated with local random motions (e.g., turbulence), 
making a thin Keplerian disk possible.
\end{abstract}


\maketitle

\section{Introduction}
\label{sec1}

Accretion onto a gravitating object is believed to be a principal energy 
source for powering many astrophysical systems, including active galactic 
nuclei (AGN), quasars, and some types of stellar binaries \cite{fra02}. 
Since accreting gases far from the central object typically have 
sufficiently large specific angular momentum compared to a particle
moving on a circular orbit around and near the central object, the 
formation of an accretion disk is unavoidable. To accrete onto the central
object and release gravitational energy, gases at large distance must 
lose a lot of angular momentum. It is usually assumed that some kind of
viscosity slowly operates in a disk to transport angular momentum outward
and dissipate energy, so that to a very good approximation in the disk
region particles move on circular orbits with a small radial velocity 
superposed.

The simplest stationary and axisymmetric model of accretion disk is a 
geometrically thin 
Keplerian disk, where it is assumed that in the disk region the gravity of 
the central object is approximately balanced by the centrifugal force so 
that disk particles move approximately on Keplerian circular orbits 
\cite{sha73,nov73,pag74}. Then, at each radius, the specific angular 
momentum and the specific energy of disk particles are given by that of a 
test particle moving on the Keplerian circular orbit at that radius. When the 
central object is a black hole, the Keplerian disk has an inner boundary at 
the marginally stable circular orbit, inside which the gravity is too 
strong to be balanced by the centrifugal force \cite{bar72,nov73,pag74}. 
The total efficiency of converting mass into energy is defined by
\begin{eqnarray}
	\varepsilon \equiv \frac{\delta {\cal E}}{\delta M} \;,
\end{eqnarray}
where $\delta {\cal E}$ is the total released energy during the process of
accretion, $\delta M$ is the corresponding accreted mass \footnote{Throughout 
the paper we use geometrized units $G = c = 1$, where $G$ is the Newtonian 
gravitational constant, $c$ is the speed of light.}. Then, the total 
efficiency of a Keplerian disk around a Kerr black hole is given by the 
difference between the specific energy at infinity ($E_\infty = 1$) and the 
specific energy on the marginally stable circular orbit ($E_{\rm ms}$)
\begin{eqnarray}
	\varepsilon_{\rm Keplerian} = 1 - E_{\rm ms} \;.
	\label{ek}
\end{eqnarray}

Thin Keplerian disks correspond to the most efficient case of releasing the
gravitational energy of an accretion flow. For a thin Keplerian disk around 
a Schwarzschild black hole, the total efficiency is $\varepsilon_{\rm 
Keplerian} \approx 0.06$. For a thin Keplerian disk around a ``canonical'' 
Kerr black hole of specific spinning angular momentum $a = 0.998 M$, where 
$M$ is the mass of the black hole, the total efficiency is $\varepsilon_{\rm 
Keplerian} \approx 0.32$ when the effect of radiation capture is considered 
\cite{tho74}. These efficiencies of converting mass into energy are indeed 
much higher than nuclear fusion, the latter is $\approx 0.007$ for hydrogen. 
The released gravitational energy in a thin Keplerian disk is assumed to be 
efficiently dissipated by the internal viscosity and instantly radiated away 
from the surfaces of the disk.

Though the model of geometrically thin and optically thick Keplerian 
accretion disks is supported by the appearance of Big Blue Bump in the 
spectra of all bright AGN (\cite{shi78,kor99,kis03} and references therein), 
it is challenged by the observational fact that most nearby galactic nuclei 
are much less active---sometimes not active at all (\cite{nar02} and 
references therein). These objects have luminosities that are lower than what 
a thin Keplerian disk model usually predicts by many orders of magnitudes. In 
order to interpret this phenomenon, different disk models are proposed where 
either the radiative efficiency is assumed to be low so that the released 
gravitational energy is converted into entropy and internal energy trapped in 
the flow and carried into the central black hole by the flow, or the mass 
accretion rate is assumed to be low and most of the released gravitational 
energy is carried away by a mass outflow 
\cite{nar94,bla99b,sto99,nar00,qua00,igu03}. 
These alternative models are summarized by Blandford \cite{bla99a} and 
Narayan \cite{nar02}. Though they predict a different disk luminosity, these
models are in common with the thin Keplerian disk in the following 
aspect: the mechanical energy (= gravitational potential energy + kinetic
energy) of the disk flow is efficiently converted into other forms of 
energy. 

In the theory aspect, the proposed disk viscosity is poorly understood. 
Investigations on the magnetohydrodynamics (MHD) in a differentially 
rotating disk suggest that magnetic fields may play the role of disk 
viscosity (\cite{bal98,bla02} and references therein). Due to the existence of 
magnetorotational instability, MHD turbulence is easily generated in a 
differentially rotating disk, which effectively transports angular momentum 
outward and drives accretion (\cite{bal91,sto96,haw96,haw00,haw02} and
references therein). However, very few models of disk radiation are based on 
these ideas, and generally angular momentum transport is treated through a 
parameterized viscous stress---the so-called $\alpha$-viscosity 
following Shakura \& Sunyaev \cite{sha73,kor99}. This is 
because of the fact that though it is well established that magnetic fields 
effectively transport angular momentum and drive accretion, it is far from 
clear how magnetic fields transport and dissipate energy in the meantime.
Numerical simulations are a powerful tool for studying the nonlinear evolution
of magnetic fields. However, numerical simulations usually have large 
numerical dissipation due to finite spatial resolutions, which makes energy 
transportation and dissipation cannot be accurately studied with current 
numerical simulations \cite{haw95,ryu95,bal98,bla02}.
Therefore, our current understanding on how magnetic fields transport 
and dissipate energy is much worse than our understanding on how magnetic 
fields transport angular momentum. 

In this paper we ask and try to answer the following question: {\it Do 
magnetic fields transport or dissipate energy as efficiently as they 
transport angular momentum?} Or, more precisely, {\it Do magnetic fields 
have to transport or dissipate a lot of energy as they efficiently 
transport angular momentum in a disk accretion flow?}

To answer this question (or, to get some insight into the answer), we use an 
analytical model to study the dynamical effects of magnetic fields in an 
accretion disk around a Kerr black hole. The disk is assumed to be stationary, 
axisymmetric, perfectly conducting, and contain a large-scale magnetic 
field~\footnote{A large-scale magnetic field in an accretion disk can arise 
in at least two ways: 1)~Inherited from the progenitor of the disk. For 
example, merger of a magnetized neutron star with a black hole will naturally 
lead to the formation of a disk with a large-scale magnetic field. 2)~Generated 
from small-scale magnetic fields through reconnection. See, e.g., C. A. Tout \& 
J. E. Pringle, Mon. Not. R. Astron. Soc. {\bf 281}, 219 (1996).}. 
Both the magnetic field and the velocity field of the disk fluid have only 
radial and azimuthal components in a small neighborhood of the disk central 
plane $\theta = \pi/2$, where $\theta$ is the polar angle from 
the symmetry axis of the black hole. In other words, near the central plane of 
the disk, each magnetic field line lies on a surface of constant $\theta$, and 
the same is true for each fluid stream line. (Similar approach is usually used 
to study the inflow/outflow problem associated with a low density magnetosphere 
around a black hole or neutron star, see 
\cite{cam86,cam86a,tak90,hir92,dai02,tak02}, and \cite{pun01} for a
review.) For simplicity, we neglect the thermal pressure in the disk by
assuming that the accretion flow is cold. Then, making use of the conservation
of rest mass, energy, and angular momentum, Maxwell's equations and dynamical 
equations are reduced to a set of one-dimensional equations with radius $r$ as 
the only variable. We will self-consistently solve these equations, to look for
the relation between accretion and the transportation of angular momentum 
and energy.

Gammie \cite{gam99} has used the model to study the flow on the equatorial 
plane in the plunging region between the inner boundary of the disk and the
horizon of the black hole. Gammie's model is further explored by Li \cite{li03} 
and tested with numerical simulations by De Villiers \& Hawley \cite{vil02}.
In this paper we extend the model into the disk region, and into a small 
neighborhood of the equatorial plane. This extension allows us to investigate 
a disk accretion flow driven by magnetic fields from a global view: from the 
subsonic far outer disk region down to the supersonic deep plunging 
region~\footnote{In this paper we use the word ``sonic'' to refer to 
``magnetosonic'' since we have assumed that the gas pressure is zero.}. We 
emphasize that such a model can only be applied to a large-scale and ordered 
magnetic field. For small-scale and chaotic
magnetic fields, the gradient of the magnetic field is important, making the
solutions invalid when even a small nonzero resistivity is introduced (see
Sec.~\ref{sec6} for detail). 

The paper is organized as follows: In Sec.~\ref{sec2} we exactly solve 
Maxwell's equations for the model outlined above. In Sec.~\ref{sec3} we 
write down the equations for the conservation of rest mass, the conservation 
of angular momentum, and the conservation of energy. We show that the solutions 
always correspond to an accretion flow with {\it constant specific energy} when 
reasonable boundary conditions at infinity are satisfied. Then we derive a 
one-dimensional radial momentum equation corresponding to the magnetized disk 
accretion flow.  In Sec.~\ref{sec4} we discuss the analytical properties of the 
radial momentum equation, the solutions on the horizon of the black hole, and 
the asymptotic solutions at infinity. We show that the fast critical point is 
the only intrinsic singularity in the radial momentum equation. In 
Sec.~\ref{sec5} we present some explicit solutions representing a magnetized 
disk flow with constant specific energy accreting from infinity onto a central 
black hole. In Sec.~\ref{sec6} we discuss the effect of finite resistivity and 
compare the evolution of large-scale magnetic fields to that of small-scale 
magnetic fields. In Sec.~\ref{sec7} we summarize the results and draw our 
conclusions. 

In Appendix~\ref{appa} we make some effort to generalize our results to more 
general and more complicated models. In Appendix~\ref{appb} we present the 
Newtonian limit of our solutions.

\section{Solutions to Maxwell's Equations}
\label{sec2}

We assume that the background spacetime is described by the metric of a
Kerr black hole of mass $M$ and specific angular momentum $a$, where
$-M \le a \le M$. In Boyer-Lindquist coordinates $(t,r,\theta,\phi)$, the 
metric of a Kerr black hole is given by \cite{mis73,wal84}
\begin{eqnarray}
        ds^2 = -\left(1-\frac{2M r}{\Sigma}\right) dt^2
                -\frac{4M a r}{\Sigma}\, \sin^2\theta\, dt d\phi
                +\frac{\Sigma}{\Delta}\, dr^2 +\Sigma\, d\theta^2
                +\frac{\Lambda\sin^2\theta}{\Sigma}\, d\phi^2\;,
    \label{gab0}
\end{eqnarray}
where
\begin{eqnarray}
        \Delta \equiv r^2-2M r+a^2\;, \hspace{1cm} \Sigma \equiv
        r^2+a^2\cos^2\theta\;,
        \hspace{1cm} \Lambda \equiv (r^2+a^2)^2-\Delta a^2 \sin^2\theta\;.
        \label{del}
\end{eqnarray}
The radius of the outer event horizon is at $r_{\rm H} = M + \left(M^2 - 
a^2\right)^{1/2}$.

As stated in the Introduction, we focus on solutions in a small neighborhood 
of the equatorial plane defined by $\cos^2\theta \ll 1$. Then, to the first 
order of $\cos\theta$ \footnote{When we say that an expression is accurate to 
the $n$-th order of $x$, we mean that the difference between the value of the
expression and the exact value is at the order of $x^{n+1}$ or higher.}, the 
metric does not depend on $\theta$ and is given by
\begin{eqnarray}
	ds^2 = -\left(1-\frac{2M}{r}\right) dt^2 -\frac{4M a}
                {r}\, dt d\phi +\frac{r^2}{\Delta}\, dr^2 +
		r^2 d\theta^2 +\frac{A}{r^2}\, d\phi^2 \;,
    \label{gab}
\end{eqnarray}
where $A \equiv \Lambda(\theta = \pi/2)$.

Assuming that the disk fluid is perfectly conducting so that magnetic field 
lines are frozen to the fluid. Then, in terms of coordinate components, the 
Maxwell equations that we need to solve are reduced to \cite{lic67,li03}
\begin{eqnarray}
        \frac{1}{\sqrt{-g}} \frac{\partial}{\partial x^\alpha}
                \left[\sqrt{-g} \left(u^\alpha B^\beta - u^\beta
                B^\alpha\right)\right] = 0 \;,
        \label{maxeq}
\end{eqnarray}
where $x^\alpha = (t,r,\theta,\phi)$, $\sqrt{-g} = r^2$ (to the first order of 
$\cos\theta$), $u^\alpha$ is the four-velocity of the fluid, and $B^\alpha$ is 
the magnetic field measured by an observer comoving with the fluid. (The 
electric field measured by an observer comoving with a perfectly conducting 
fluid is zero.) 

We assume that in the small neighborhood of the equatorial plane we have
\begin{eqnarray}
        u^\theta = 0 \;, \hspace{1cm} B^\theta =0 \;,
	\label{bth}
\end{eqnarray}
i.e., each line of velocity or magnetic field lies on a surface of $\theta 
= \mbox{constant}$. Equation~(\ref{bth}) implies that 
\begin{eqnarray}
        \frac{\partial u^\theta}{\partial \theta} = 0 \;,
	        \hspace{1cm}
	\frac{\partial B^\theta}{\partial \theta} = 0 \;,
	        \label{dbth}
\end{eqnarray}
in the same neighborhood. Then, for stationary and axisymmetric solutions 
with $\partial /\partial t = \partial /\partial \phi = 0$, the Maxwell
equation~(\ref{maxeq}) is reduced to
\begin{eqnarray}
	\frac{\partial}{\partial r} \left[r^2 (u^r B^\beta 
		- u^\beta B^r)\right] = 0 \;.
	\label{eq1}
\end{eqnarray} 

By definition, $B^a$ satisfies $u_a B^a = 0$. Then, making use of $u_au^a = -1$, 
we can solve for $B^r$, $B^\phi$, and $B^t$ from Eq.~(\ref{eq1}). The results are
\begin{eqnarray}
	B^r &=& \frac{1}{r^2} (-C_0 u_t + \Psi u_\phi) \;, 
		\label{br}\\
	B^\phi &=& \frac{1}{r^2 u^r} [-C_0 u_t u^\phi + (1+ 
                u_\phi u^\phi)\Psi] \;, \label{bf}\\
	B^t &=& \frac{1}{r^2 u^r} [-(1+u_t u^t) C_0 +
                u_\phi u^t\Psi] \;,
	        \label{bt}
\end{eqnarray}
where $C_0$ and $\Psi$ are constants. 

The electromagnetic field tensor corresponding the solutions given by 
Eqs.~(\ref{br})--(\ref{bt}) is
\begin{eqnarray}
	F_{ab} = \epsilon_{abcd} u^c B^d = -2 \Psi\, dt_{[a} d\theta_{b]} +
		\frac{2}{u^r}\left(C_0 u^\phi + \Psi u^t\right) 
		dr_{[a} d\theta_{b]} + 2 C_0\, d\theta_{[a} d\phi_{b]} 
		\;,
	\label{fab}
\end{eqnarray}
where $\epsilon_{abcd}$ is the totally antisymmetric tensor of the volume 
element that is associated with the metric \cite{wal84}, the brackets 
``[~]'' denote antisymmetrization. Gammie \cite{gam99} obtained the same 
solution for $F_{ab}$ by directly solving the Maxwell equation $\nabla_{[a} 
F_{bc]} = 0$.

It is useful to have a look at the magnetic field and the electric field 
measured by an observer in the locally nonrotating frame (LNRF observer, see 
\cite{bar72}) in the equatorial plane, whose four-velocity is $U^a = \chi^{-1} 
\left[\left(\partial/\partial t\right)^a + \omega \left(\partial/\partial
\phi\right)^a\right]$,
where $\chi \equiv \left(r^2\Delta/A\right)^{1/2}$ and $\omega \equiv 2 Mar/A$
are respectively the lapse function (i.e., the redshift factor) and the
frame dragging angular velocity. From the electromagnetic field tensor in 
Eq.~(\ref{fab}), we can calculate the magnetic field and the electric 
field measured by a LNRF observer
\begin{eqnarray}
	B^{\prime a} &\equiv& -\frac{1}{2} \epsilon^a_{~bcd} U^b F^{cd}
		= \frac{\chi C_0}{r^2} \left(\frac{\partial}
		{\partial r}\right)^a + \frac{\chi}{r^2 u^r}
		\left(C_0 u^\phi + \Psi u^t\right) 
	        \left(\frac{\partial}{\partial\phi}\right)^a \;,
	\label{mfield} \\
	E^{\prime a} &\equiv& F^a_{~b} U^b = \frac{1}{\chi r^2}(C_0 
		\omega + \Psi) \left(\frac{\partial}{\partial\theta}
		\right)^a \;.
	\label{efield}
\end{eqnarray}

Therefore, the constant $C_0$ measures the flux of the radial magnetic field,
and the constant $\Psi$ is related to the electric field measured by a LNRF
observer and thus measures the direction of the magnetic field relative to 
the fluid motion \cite{li03}.

Equation~(\ref{br}) can be rewritten as $\Psi = u_\phi^{-1}(r^2 B^r + C_0 
u_t)$. For a Keplerian disk we have $u_t = -1$ and $u_\phi \propto r^{1/2}$ 
as $r\rightarrow \infty$. If generally we assume that $u_t$ and
$r^2 B^r$ are bounded but $u_\phi$ is unbounded as $r\rightarrow \infty$, 
then we must have 
\begin{eqnarray}
        \Psi = 0 \;, \label{limp}\hspace{0.6cm} \mbox{i.e.,}\hspace{0.6cm}
	   \frac{B^r}{B^\phi} = \frac{u^r}{u^\phi} \;. \label{parel}
\end{eqnarray}
We note that it is not obvious that $r^2 B^r$ must be finite, since the 
magnetic flux across a circle of a radius $r$ is $\propto r^2 B^{\prime r}/
\chi$, not $r^2 B^r$. Stronger justifications for the validity of 
Eq.~(\ref{parel}) are given in the next section where dynamical equations are 
discussed. 

From Eqs.~(\ref{br}) and (\ref{bf}) we have
\begin{eqnarray}
	k\equiv 1 -\frac{B^\phi u^r}{B^r u^\phi} = -\frac{\Psi}{\left(
		-C_0 u_t + \Psi u_\phi\right) u^\phi} \;.
	\label{upp}
\end{eqnarray}
Because of the appearance of $u^\phi$ in the denominator on the right-hand
side, which generally quickly decreases with radius (e.g.,
$u^\phi/u^t \propto r^{-3/2}$ for a Keplerian disk), $|k|$ grows quickly as 
$r$ increases, if $\Psi\neq 0$. This means that, if at a large radius $k$ is 
$\sim 1$, then $k$ quickly approaches zero as $r$ decreases. 

As an example, assuming that at $r = r_0 \gg r_{\rm H}$ we have $k = k_0 \sim 
1$ (which is true if $B^r \sim r B^\phi$ and $|u^r/r u^\phi|\ll 1$ at $r = 
r_0$), then by Eq.~(\ref{upp}) we have
\begin{eqnarray}
	\Omega_\Psi \equiv -\frac{\Psi}{C_0} = \left.-\frac{k_0 u_t u^\phi}{1 
	     + k_0 u_\phi u^\phi}\right|_{r = r_0} \approx k_0 \Omega_0 \;,
	\label{opsi}
\end{eqnarray}
where $\Omega_0 \equiv \left(u^\phi/u^t\right)_{r = r_0}$ is the angular 
velocity of the disk at $r = r_0\,$. In the last step we have used the fact 
that $u_t \approx -u^t\approx -1$ and $u_\phi u^\phi \sim M/r \ll 1$ at $r = 
r_0$\,. Substitute Eq.~(\ref{opsi}) into Eq.~(\ref{upp}), we then obtain
\begin{eqnarray}
	k \approx -\frac{k_0 \Omega_0}{\left(u_t + k_0 \Omega_0 u_\phi
		\right)u^\phi} 
	\nonumber
\end{eqnarray}
at any radius $r$. For $r\ll r_0$ in a Keplerian-type disk we have
\begin{eqnarray}
	|k| \approx \left|\frac{k_0 \Omega_0}{u_t u^\phi}\right| 
		\sim\left(\frac{r}{r_0}\right)^{3/2} \ll 1\;,
	\label{ka1}
\end{eqnarray}
since $u_0^\phi u_\phi \sim (M/r_0)(r/r_0)^{1/2} \ll 1$ and $u_t \sim -1$.

When Eq.~(\ref{parel}) is satisfied, from Eq.~(\ref{efield}) we have $E^{\prime 
a}= \left(C_0 \omega/\chi r^2\right) (\partial/\partial\theta)^a$. Then, for
a Schwarzschild black hole (i.e., $a = \omega = 0$), we have $E^{\prime a} =0$,
i.e., the magnetic field is parallel to the velocity field as seen by a local
static observer (equivalent to a LNRF observer for a Kerr black hole). For a Kerr
black hole with $a\neq 0$, the electric field measured by a LNRF observer is
nonzero. However, the electric field decays quickly with increasing $r$:
$E^\prime_a E^{\prime a} \propto r^{-8}$ for $r\gg r_{\rm H}$. Therefore, for a 
Kerr black hole, to a good approximation the magnetic field is also parallel to
the velocity field in a LNRF, unless in the region close to the horizon of the
black hole. Therefore, when Eq.~(\ref{parel}) is satisfied, we say that the 
magnetic field is parallel to the velocity field; or equivalently, the magnetic 
field lines are aligned with the fluid motion. But it must be kept in mind that 
this statement is not very accurate for the case of a Kerr black hole according 
to the above discussion. 

Equation~(\ref{ka1}) indicates that after the disk flow settles into the 
stationary and axisymmetric state the magnetic field and the velocity field
tend to be aligned with each other if they are not so initially.

In the next section we will see that a magnetic field satisfying 
Eq.~(\ref{parel}) has a very interesting feature: it does not change the 
specific energy of fluid particles though it affects the specific angular 
momentum. Physically this is caused by the fact that the electromagnetic force 
produced by the magnetic field satisfying Eq.~(\ref{parel}) is always 
perpendicular to the velocity of the fluid.

\section{Derivation of the Radial Momentum Equation}
\label{sec3}

For a cold and perfectly conducting magnetized fluid the total stress-energy 
tensor is given by $T^{ab} = \rho_{\rm m} u^a u^b + T_{\rm EM}^{ab}$, where 
$\rho_{\rm m}$ is the mass-energy density of the fluid matter as measured by 
an observer comoving with the fluid, $T_{\rm EM}^{ab}$ is the stress-energy 
tensor of the magnetic field \cite{lic67,li03}
\begin{eqnarray}
        T_{\rm EM}^{ab} = \frac{1}{4\pi} B^2 u^a u^b + \frac{1}{8\pi}
                      B^2 g^{ab} - \frac{1}{4\pi} B^a B^b \;.
        \label{tab}
\end{eqnarray}
By Einstein's equation, the dynamical equations of the magnetized fluid are 
given by \cite{mis73,wal84}
\begin{eqnarray}
	\nabla_a T^{ab} = 0 \;.
	\label{dtab}
\end{eqnarray}

\subsection{Conservation laws}
\label{sec3.1}

A number of conservation laws can be derived from Eq.~(\ref{dtab}) 
\cite{lic67,bek78,wal84,li03}. The contraction of $u_b$ with Eq.~(\ref{dtab}) 
leads to the {\it conservation of rest mass} (also called the {\it continuity 
equation}): $\nabla_a \left(\rho_{\rm m} u^a\right) = 0$\,. The contraction
of $(\partial/\partial \phi)^a$ with Eq.~(\ref{dtab}) leads to the {\it 
conservation of angular momentum}: $\nabla_a T_{\phi}^{~a} = 0$\,. The 
contraction of $(\partial/\partial t)^a$ with Eq.~(\ref{dtab}) leads to 
the {\it conservation of energy}: $\nabla_a T_{t}^{~a} = 0$\,. [Note that
$(\partial/\partial t)^a$ and $(\partial/\partial \phi)^a$ are Killing vectors 
of a Kerr spacetime.]

Because of Eqs.~(\ref{bth}) and (\ref{dbth}), for stationary and axisymmetric 
solutions in the small neighborhood of the disk central plane, the above 
conservation equations are simplified to one-dimensional ordinary differential 
equations \cite{li03}
\begin{eqnarray}
	\frac{d}{dr}\left(r^2 \rho_{\rm m} u^r\right) &=& 0 \;;
	\label{mass} \\
	\frac{d}{dr}\left[r^2 \left(\rho_{\rm m} u_\phi u^r + 
		T_{{\rm EM},\phi}^{~~~~~r}\right)\right] &=& 0 \;;
	\label{ang} \\
	\frac{d}{dr}\left[r^2 \left(\rho_{\rm m} u_t u^r + 
		T_{{\rm EM},t}^{~~~~~r}\right)\right] &=& 0 \;;
	\label{ener}
\end{eqnarray}
respectively.

Substituting the solutions~(\ref{br})--(\ref{bt}) into Eq.~(\ref{tab}), we 
obtain
\begin{eqnarray}
        T_{{\rm EM},\phi}^{~~~~~r} &=& \frac{1}{4\pi} \left(B^2 u_\phi u^r
		- B_\phi B^r\right) = - \frac{C_0 \Delta}{4\pi r^4 u^r}
		\left(C_0 u^\phi + \Psi u^t\right)\;, \hspace{0.6cm} 
		\label{tfr} \\
        T_{{\rm EM},t}^{~~~~~r} &=& \frac{1}{4\pi} \left(B^2 u_t u^r
		- B_t B^r\right) = - \Omega_\Psi T_{{\rm EM},
		\phi}^{~~~~~r} \;,
        \label{ttr}
\end{eqnarray}
where $\Omega_\Psi$ is defined by Eq.~(\ref{opsi}). (Though based on the 
arguments in Sec.~\ref{sec2} we expect $\Psi = 0$ in a stationary state, in 
this section we keep $\Psi$ in the derived formulas to make the formulas
as general as possible.)

The solution to Eq.~(\ref{mass}) is
\begin{eqnarray}
	\rho_{\rm m} = \frac{F_{\rm m}}{4 \pi r^2 u^r} \;,
        \label{rhm}
\end{eqnarray}
where $F_{\rm m}$ is the constant of radial mass flux. From 
Eqs.~(\ref{br})--(\ref{bt}), at large radii where $r \gg M$, $u_t \approx 
- u^t \approx -1$, $u^r \approx 0$, $u^\phi \approx 0$, and $u_\phi$ is 
unbounded, we have $B^2 = B_a B^a \approx \Psi^2/r^2 \left(u^r\right)^2$,
if $\Psi \neq 0$ and $u_\phi^2/r^2 \ll 1$ (e.g., $u_\phi^2/r^2 
\approx M/r$ for a Keplerian disk). Then
\begin{eqnarray}
	\frac{B^2}{4\pi \rho_{\rm m}} \approx \frac{\Psi^2}
		{F_{\rm m} u^r} \propto \frac{1}{-u^r} \;,
	\label{b2ro}
\end{eqnarray}
as $r\rightarrow \infty$. Equation~(\ref{b2ro}) implies that if $\Psi\neq
0$, at sufficiently large radii the magnetic field must be so strong that
$B^2/4\pi \gg \rho_{\rm m}$. In order to prevent this to happen, $\Psi$
must be zero. This provides a strong support for Eq.~(\ref{parel}).

Making use of Eqs.~(\ref{tfr})--(\ref{rhm}), we can integrate Eqs.~(\ref{ang}) 
and (\ref{ener}) to obtain
\begin{eqnarray}
	u_\phi + \frac{c_0^2 \Delta}{r^2 u^r} \left(u^\phi 
		-\Omega_\Psi u^t\right) = -f_{\rm L} \;, 
		\label{fl} \\
        u_t - \Omega_\Psi \frac{c_0^2 \Delta}{r^2 u^r} \left(u^\phi -
		\Omega_\Psi u^t\right) = f_{\rm E} \;, 
		\label{fe}
\end{eqnarray}
where
\begin{eqnarray}
        c_0 \equiv \frac{C_0}{\sqrt{-F_{\rm m}}} \;, \hspace{1cm}
        f_{\rm L} \equiv \frac{F_{\rm L}}{- F_{\rm m}} \;, \hspace{1cm}
        f_{\rm E} \equiv \frac{F_{\rm E}}{- F_{\rm m}} \;,
	\label{c0etc}
\end{eqnarray}
$F_{\rm L} = 4\pi r^2 T_\phi^{~r}$ and $F_{\rm E} = -4\pi r^2 T_t^{~r}$ 
are the constants of radial angular momentum flux and radial energy flux, 
respectively. The dynamical equations~(\ref{fl}) and (\ref{fe}) do 
not depend on the sign of $c_0$. Without loss of generality, henceforth we 
assume that $c_0 >0\,$.

From Eqs.~(\ref{fl}) and (\ref{fe}) we have
\begin{eqnarray}
	E - \Omega_\Psi L = -f_{\rm E} +\Omega_\Psi f_{\rm L} \;,
	\label{utuf}
\end{eqnarray}
where $E = - u_t$ is the specific energy, $L = u_\phi$ is the specific angular
momentum of fluid particles. Since $f_{\rm E}$, $f_{\rm L}$, and $\Omega_\Psi$ 
are constants, Eq.~(\ref{utuf}) immediately implies that the specific 
energy $E$ must be constant when $\Psi = 0$ (i.e., $\Omega_\Psi = 0$), and 
the specific angular momentum $L$ must be constant when $C_0 = 0$ but $\Psi
\neq 0$ (i.e., $\left|\Omega_\Psi\right| = \infty$).

The variation of Eq.~(\ref{utuf}) leads to
\begin{eqnarray}
	\delta E = \Omega_\Psi \delta L \;.
	\label{dele}
\end{eqnarray}
Obviously, $\Omega_\Psi$ cannot be negative if we want $\delta E/\delta
L \ge 0\,$, i.e., angular momentum and energy propagate in the same direction.

If the outer boundary of the disk flow is at $r = \infty$ and the specific 
angular 
momentum $L$ is unbounded at $r = \infty$, then we have $\delta L = L(r = 
\infty) - L(r) = \infty$ for any finite $r$. Generally $E$ is positive and 
bounded at $r = \infty$ and $E$ must be positive outside the ergosphere of a 
Kerr black hole. Hence, $\delta E = E(r = \infty) - E(r)$ must be finite
for $r> 2M$. Then, by Eq.~(\ref{dele}), this can be true if and only if 
$\Omega_\Psi = 0$, which provides an additional support for Eq.~(\ref{parel}) 
from the consideration of energy conservation.

If the outer boundary of the disk flow is at a finite radius $r_0\gg 
r_{\rm H}$ and $\Omega_\Psi$ is given by Eq.~(\ref{opsi}), then, by
Eq.~(\ref{dele}), for any $r\ll r_0$ we have
\begin{eqnarray}
	\delta E = E_0 - E(r) \approx k_0 \Omega_0 L_0 \;,
\end{eqnarray}
where $E_0 \equiv E(r = r_0)$, $L_0 \equiv L(r = r_0)$, and we have assumed 
that $L_0 \gg L(r)$. Notice that the value of $\delta E$ is determined by
the boundary condition at the {\it outer} boundary of the disk when $r \ll r_0$, 
which is in contrast with the case of a thin Keplerian disk where the total 
energy variation is determined by the boundary condition at the {\it inner}
boundary of the disk. This is caused by the fact that in the present model the 
variation in energy can only happen in a region close to the outer boundary, 
since the magnetic field and the velocity of the fluid quickly get aligned with 
each other in the region of $r\ll r_0$.   

\subsection{Dynamical equilibrium in the $\theta$-direction}
\label{sec3.2}

Having solved the equations for the conservation of rest mass, angular 
momentum, and energy on each surface of $\theta = \mbox{constant}$ near the 
equatorial plane, we need to check if the dynamical equilibrium in the 
$\theta$-direction is maintained. 

To do so, let us define a vector
\begin{eqnarray}
	\left(\frac{\partial}{\partial z}\right)^a \equiv -
		\frac{1}{\sin\theta} \left(\frac{\partial}{\partial 
		\theta}\right)^a \;, 
	\hspace{1cm}
	z \equiv \cos\theta \;.
	\label{dza}
\end{eqnarray}
Since $T^{ab} = T^{ba}$, the contraction of $(\partial/\partial z)_b$ with 
Eq.~(\ref{dtab}) leads to
\begin{eqnarray}
	\nabla_a \left[\left(\frac{\partial}{\partial z}\right)_b 
		T^{ab}\right]= T^{ab} \nabla_{(a} \left(\frac{\partial}
		{\partial z}\right)_{b)} \;,
	\label{dtz}
\end{eqnarray}
where the braces ``(~)'' in the subscripts denote symmetrization of tensor.
Equation~(\ref{dtz}) governs the dynamical equilibrium in the $\theta$-direction. 

It is easy to check that
\begin{eqnarray}
	\nabla_{(a} \left(\frac{\partial}{\partial z}\right)_{b)}
		= (\mbox{terms} \propto z) + {\cal O}(z^3) \;, 
	\label{dzz}
\end{eqnarray}
near the equatorial plane ($z = 0$). The notation ${\cal O}(x^n)$ denotes
terms of order $x^n$ or higher. Since $T^{ab}$ is regular and nonzero on 
the equatorial plane, from Eqs.~(\ref{dtz}) and (\ref{dzz}) we have
\begin{eqnarray}
	\nabla_a \left[\left(\frac{\partial}{\partial z}\right)_b 
		T^{ab}\right] \propto z
	\label{dtz2}
\end{eqnarray}
for small $|z|$.

For our model, $T_z^{~r} = 0$ and $T_z^{~z} = B^2/8\pi$, so we have 
\begin{eqnarray}
	\nabla_a \left[\left(\frac{\partial}{\partial z}\right)_b 
		T^{ab}\right]  = \frac{1}{\Sigma}\frac{\partial}
		{\partial z} \left(\Sigma\,\frac{B^2}{8\pi}\right) \;,
	\label{dtz3}
\end{eqnarray}
where $\Sigma$ is defined by Eq.~(\ref{del}). Then, by Eqs.~(\ref{dtz2}) and
(\ref{dtz3}) we have
\begin{eqnarray}
	B^2 = (\mbox{terms independent of $\theta$}) + 
		{\cal O}(\cos^2\theta) \;
	\label{b2z}
\end{eqnarray}
for $\cos^2\theta\ll 1$\,. Equation~(\ref{b2z}) indicates that to the {\it 
first} order of $\cos\theta$, the dynamical equilibrium in the 
$\theta$-direction is guaranteed though the variable $\theta$ does not appear 
in our solutions. Indeed, the solutions are exact on the equatorial plane 
$\theta = \pi/2\,$.

\subsection{The radial momentum equation}
\label{sec3.3}

Using $u_\phi = g_{\phi t} u^t + g_{\phi\phi} u^\phi$ and $u_t = g_{tt} u^t 
+ g_{t \phi} u^\phi$, we can solve for $u^t$ and $u^\phi$ from 
Eqs.~(\ref{fl}) and (\ref{fe}). The results are
\begin{eqnarray}
	u^t &=& - \frac{\frac{A}{r^2\Delta}\left(f_{\rm E} -\omega
		f_{\rm L}\right) + \frac{c_0^2}{r^2 u^r}\left( 
		f_{\rm E} -\Omega_\Psi  f_{\rm L}
		\right)}{1+ \frac{c_0^2}{r^2 u^r}\left[\chi^2 - 
		\frac{A}{r^2}\left(\omega-\Omega_\Psi\right)^2\right]}\;,
	\label{utu} \\
	u^\phi &=& -\frac{\frac{1}{\Delta}\left[\frac{A}{r^2}\omega 
		f_{\rm E} + 
		\left(1-\frac{2M}{r}\right) f_{\rm L}\right]+\Omega_\Psi  
		\frac{c_0^2}{r^2 u^r}\left(f_{\rm E} -\Omega_\Psi 
		f_{\rm L}\right)}{1+ \frac{c_0^2}{r^2 u^r}\left[\chi^2 - 
		\frac{A}{r^2}\left(\omega-\Omega_\Psi\right)^2\right]}\;.
	\label{uphiu}
\end{eqnarray}
Here and in some of the following equations we substitute the frame 
dragging frequency $\omega$ for the specific angular momentum $a$ to 
simplify expressions. 

The corresponding covariant components are
\begin{eqnarray}
	-E &=& u_t ~=~ \frac{f_{\rm E} + \frac{c_0^2}{r^2 u^r}
		\left(f_{\rm E} - \Omega_\Psi f_{\rm L}\right)
		\left(1-\frac{2M}{r} +\frac{A}{r^2}\omega \Omega_\Psi
		\right)}{1+ \frac{c_0^2}{r^2 u^r}\left[\chi^2 - 
		\frac{A}{r^2}\left(\omega-\Omega_\Psi
		\right)^2\right]}\;,
	\label{utd} \\
	L &=& u_\phi ~=~ \frac{-f_{\rm L} + \frac{A}{r^2}\frac{c_0^2}{r^2 
		u^r}\left(f_{\rm E} - \Omega_\Psi f_{\rm L}\right)
		\left(\omega- \Omega_\Psi\right)}
		{1+ \frac{c_0^2}{r^2 u^r}\left[\chi^2 - 
		\frac{A}{r^2}\left(\omega-\Omega_\Psi\right)^2\right]}
		\;.
	\label{uphid}
\end{eqnarray}
The covariant radial velocity is $u_r = g_{rr} u^r = r^2 u^r/\Delta$\,.

Having expressed all components of $u^a$ and $u_a$ in terms of $u^r$,
we are ready to write down the {\it radial momentum equation} for $u^r$ 
by making use of $u_a u^a = -1$. The result is
\begin{eqnarray}
	\left(r u^r\right)^2 &=& -\Delta\left[1 - \frac{
		\left(f_{\rm E} -\Omega_\Psi f_{\rm L}\right)^2}
		{\chi^2 - \frac{A}{r^2}\left(\omega-\Omega_\Psi
		\right)^2}\right]  \nonumber\\ 
		&&-\frac{1}{\chi^2 - \frac{A}{r^2}\left(\omega-
		\Omega_\Psi\right)^2} \left\{\frac{\frac{A}{r^2}
		\left(\omega - \Omega_\Psi\right)f_{\rm E} +\left(1 
		-\frac{2M}{r} + \frac{A}{r^2}\omega\Omega_\Psi\right)
		f_{\rm L}}{1+ \frac{c_0^2}{r^2 
		u^r}\left[\chi^2 - \frac{A}{r^2}\left(\omega-
		\Omega_\Psi\right)^2\right]}\right\}^2 \;.
	\label{floweq}
\end{eqnarray}

The radial momentum equation~(\ref{floweq}) is a nontrivial algebraic equation: 
it has multiple roots and singularities. Any physical solution must smoothly 
pass these singularities, which gives rise to strong constraints on physically 
allowable integral constants. As we will see in the next section, among the 
four integral constants appearing in Eq.~(\ref{floweq}), $c_0$, $\Omega_\Psi$, 
$f_{\rm E}$, and $f_{\rm L}$, only three of them are independent for physical 
solutions.

\section{Analysis on the Radial Momentum Equation}
\label{sec4}

In order to correctly solve the radial momentum equation, we must first 
understand the analytical features and the asymptotic behaviors of the 
solutions. In this section we analyse the singularities in the radial 
momentum equation (Sec.~\ref{sec4.1}), and study the solutions on the horizon 
of the black hole (Sec.~\ref{sec4.2}) and at infinity (Sec.~\ref{sec4.3}).

\subsection{Critical points}
\label{sec4.1}

Though $\left[\chi^2 - \left(A/r^2\right)\left(\omega-\Omega_\Psi
\right)^2 \right]$ appears in the denominators on the right-hand side of 
Eq.~(\ref{floweq}), it does not represent a singularity of the equation. Indeed, 
the factor disappears from the denominators if we expand the second 
term on the right-hand side of Eq.~(\ref{floweq}), then combine with the 
first term. However, the factor
\begin{eqnarray}
	\left\{1+ \frac{c_0^2}{r^2 u^r}\left[\chi^2 - \frac{A}{r^2}
		\left(\omega-\Omega_\Psi\right)^2\right]\right\} \;,
	\label{fac1}
\end{eqnarray}
which also appears in the denominators on the right-hand side of 
Eq.~(\ref{floweq}), may represent a singularity of the equation.

Introducing the relativistic Alfv\'{e}n four-velocity $c_{\rm A}^{~\,a} \equiv 
B^a/\sqrt{4\pi\rho_{\rm m} +B^2}$, which satisfies $c_{{\rm A}a} c_{\rm 
A}^{~\,a} < 1$, we have
\begin{eqnarray}
	1+ \frac{c_0^2}{r^2 u^r}\left[\chi^2 - \frac{A}{r^2}
		\left(\omega-\Omega_\Psi\right)^2\right] =
		\frac{1}{1 - c_{\rm A}^2}
                \left( 1- \frac{c_{{\rm A}r} c_{\rm A}^{~\,r}}
                {u_r u^r}\right) \;.
        \label{crur}
\end{eqnarray}
Hence, the factor in Eq.~(\ref{fac1}) represents a singularity at the {\it 
Alfv\'{e}n critical point} defined by 
\begin{eqnarray}
	u_ru^r - c_{{\rm A}r} c_{\rm A}^{~\,r} = 0 \;,
	\hspace{1cm}
	\mbox{at $r = r_{\rm A}$} \;.
	\label{ra}
\end{eqnarray}

If we define a generation function 
\begin{eqnarray}
	F\left(r,u^r\right) &\equiv&
		\left\{1+ \frac{c_0^2}{r^2 u^r}\left[\chi^2 - 
		\frac{A}{r^2}\left(\omega-\Omega_\Psi\right)^2
		\right]\right\}^2\,\left[\frac{r^2}{\Delta}\left(u^r
		\right)^2 + 1- \frac{\left(f_{\rm E} -\Omega_\Psi 
		f_{\rm L}\right)^2}{\chi^2 - \frac{A}{r^2}
		\left(\omega-\Omega_\Psi\right)^2}\right]
		\nonumber\\
		&& +\; \frac{\left[
		\frac{A}{r^2}\left(\omega-\Omega_\Psi\right)f_{\rm E}
		+\left(1 -\frac{2M}{r} + \frac{A}{r^2}\omega\Omega_\Psi\right)
		f_{\rm L}\right]^2}{\Delta\left[\chi^2 - \frac{A}{r^2}
		\left(\omega-\Omega_\Psi\right)^2\right]} \;,
	\label{funcf}
\end{eqnarray}
then the radial momentum equation~(\ref{floweq}) is given by $F(r,u^r) = 0$. 
The corresponding differential equation for $u^r$ can be derived from
\begin{eqnarray}
	\frac{\partial F}{\partial u^r} \frac{du^r}{dr}
		+ \frac{\partial F}{\partial r} = 0 \;.
	\label{deq}
\end{eqnarray}

From Eq.~(\ref{funcf}), we have
\begin{eqnarray}
	\frac{\partial F}{\partial u^r} &=& \frac{2}{u^r}\left\{1+ 
		\frac{c_0^2}{r^2 u^r}\left[\chi^2 - \frac{A}{r^2}
		\left(\omega-\Omega_\Psi\right)^2\right]\right\} 
		\nonumber\\
		&&\times\left\{\frac{r^2}{\Delta}\left(u^r\right)^2 - 
		\frac{c_0^2}{r^2 u^r}\left[\chi^2 - \frac{A}{r^2}
		\left(\omega-\Omega_\Psi\right)^2-\left(f_{\rm E} 
		- \Omega_\Psi f_{\rm L}\right)^2\right]\right\}\;.
	\label{dfu}
\end{eqnarray}

Since $B^2 = 4\pi \rho_{\rm m} c_{\rm A}^2/\left(1 - c_{\rm A}^2\right)$ and
$B_r B^r = 4\pi \rho_{\rm m} c_{{\rm A}r}c_{\rm A}^{~\,r}/\left(1 - c_{\rm 
A}^2\right)$, by Eqs.~(\ref{br}) and (\ref{utuf}) we have
\begin{eqnarray}
	\frac{c_0^2}{r^2 u^r}\left[\chi^2 - \frac{A}{r^2}
		\left(\omega-\Omega_\Psi\right)^2-\left(f_{\rm E} 
		- \Omega_\Psi f_{\rm L}\right)^2\right] =
		\frac{c_{\rm A}^2}{1 - c_{\rm A}^2} \;.
        \label{caa}
\end{eqnarray}
Substituting Eqs.~(\ref{crur}) and (\ref{caa}) into 
Eq.~(\ref{dfu}), we obtain
\begin{eqnarray}
	\frac{\partial F}{\partial u^r} = \frac{2}{u^r\left(1 - 
		c_{\rm A}^2\right)}\left( 1- \frac{c_{{\rm A}r} c_{\rm 
		A}^{~\,r}}{u_r u^r}\right)\left(u_r u^r - 
		\frac{c_{\rm A}^2}{1 - c_{\rm A}^2}\right) \;.
	\label{dfu1}
\end{eqnarray}

For a regular and stationary flow solution, $u^r$ cannot be zero at a finite 
radius since otherwise the mass density $\rho_{\rm m}$ will diverge there. 
Therefore, Eq.~(\ref{dfu1}) implies that there are two singularities 
in the differential
radial momentum equation: one is at the Alfv\'{e}n critical point defined 
by Eq.~(\ref{ra}), the other is at the {\it fast critical point} 
defined by \footnote{Since the gas pressure is assumed to be zero, the {\it 
slow critical point}, which appears in a general magnetized wind problem, 
corresponds to $u^r = 0$ here so must be located at $r = \infty$ and is not 
relevant.}
\begin{eqnarray}
	u_r u^r =\frac{c_{\rm A}^2}{1 - c_{\rm A}^2} \;,
	\hspace{1cm}
	\mbox{at $r = r_{\rm f}$} \;.
	\label{rf}
\end{eqnarray}

In a general magnetized wind theory, the Alfv\'{e}n critical point is not 
an X-type singularity and it does not impose any additional conditions on 
the solution for $u^r$ except setting the integral constants 
\cite{web67,lam99}. For the model studied here, in fact the Alfv\'{e}n 
critical point is not a singularity at all, after the integral radial 
momentum equation (\ref{floweq}) is taken into account. To see this, notice 
that
\begin{eqnarray}
	\frac{\partial F}{\partial r} &=&
		\left\{1+ \frac{c_0^2}{r^2 u^r}\left[\chi^2 - 
		\frac{A}{r^2}\left(\omega-\Omega_\Psi\right)^2
		\right]\right\} [...] 
		\nonumber\\
		&&+\left[\frac{A}{r^2}\left(\omega-\Omega_\Psi\right)
		f_{\rm E}
		+\left(1 -\frac{2M}{r} + \frac{A}{r^2}\omega\Omega_\Psi
		\right)f_{\rm L}\right] [...] \nonumber\\
		&\propto&
		\left\{1+ \frac{c_0^2}{r^2 u^r}\left[\chi^2 - 
		\frac{A}{r^2}\left(\omega-\Omega_\Psi\right)^2
		\right]\right\} \;,
	\nonumber
\end{eqnarray}
where in the last step Eq.~(\ref{floweq}) has been substituted. Since
the factor in Eq.~(\ref{fac1}) appears in both $\partial F/\partial u^r$ and
$\partial F/\partial r$, it disappears from the final form of the differential 
radial momentum equation~(\ref{deq}). Then, only the fast critical 
point, which is defined by Eq.~(\ref{rf}), is a true singularity in the 
radial momentum equation.

Any physical solution must smoothly pass the singular fast critical point. So, 
by Eq.~(\ref{deq}), any physical solution must satisfy
\begin{eqnarray}
	\left.\frac{\partial F}{\partial u^r}\right|_{r = r_{\rm f}} 
		= 0 \;,
	\hspace{1cm}
	\left.\frac{\partial F}{\partial r}\right|_{r = r_{\rm f}} 
		= 0 
	\label{rcuc}
\end{eqnarray}
simultaneously. Equation~(\ref{rcuc}) implies that the fast critical point 
corresponds to an extremum of the generation function---indeed it is a saddle 
point of $F$ as we will see in Sec.~\ref{sec5}.

Equation~(\ref{rcuc}) together with $F(r,u^r) = 0$ allows us to solve 
for $r_{\rm f}$, $u_{\rm f}^{r} \equiv u^r(r = r_{\rm f})$, and $f_{\rm 
L}$ for given $c_0$, $\Omega_\Psi$, and 
$f_{\rm E}$. This implies that among the four integral constants $C_0$, 
$\Omega_\Psi$, $f_{\rm E}$, and $f_{\rm L}$ only three are independent.

\subsection{Solutions on the horizon of the black hole}
\label{sec4.2}

On the horizon of the black hole, where $r = r_{\rm H}$, we have $\Delta = 
\chi =0$, $\omega = \Omega_{\rm H}$, and $A = 4 M^2 r_{\rm H}^2$, where 
$\Omega_{\rm H} \equiv a/(2 M r_{\rm H})$ is the angular velocity of the 
horizon. Then, when $\Omega_\Psi\neq \Omega_{\rm H}$, at $r = r_{\rm H}$ we 
can drop 
the first term on the right-hand side of Eq.~(\ref{floweq}) and solve 
for $u_{\rm H}^r \equiv u^r (r = r_{\rm H})$. The result is
\begin{eqnarray}
	u_{\rm H}^r = \frac{2M}{r_{\rm H}}\left(f_{\rm E} -
		\Omega_{\rm H} f_{\rm L}\right) + \left(\frac{2M}
		{r_{\rm H}}\right)^2 c_0^2 \left(\Omega_{\rm H}
		- \Omega_\Psi\right)^2\;,
	\label{urh}
\end{eqnarray}
where we have used the inequality $f_{\rm E} - \Omega_{\rm H} f_{\rm L} <0$,
which comes from the requirement that on the horizon of the black hole energy 
must flow into the horizon locally.

When $\Omega_\Psi = \Omega_{\rm H}$, it can be checked that as $\delta r
\equiv r-r_{\rm H}\rightarrow 0$ we have $\chi^2 \propto \delta r$, $\omega-
\Omega_{\rm H} \propto \delta r$, $1- 2M/r + A\omega\Omega_H/r^2\propto \delta 
r$, and $\Delta/\chi^2 = A/r^2 = 4 M^2$\,.
Hence,  when $\Omega_\Psi = \Omega_{\rm H}$, on the horizon of the black 
hole we can drop the second term on the right-hand side of 
Eq.~(\ref{floweq}). Then we obtain $u_{\rm H}^r$, which is just given 
by Eq.~(\ref{urh}) with $\Omega_\Psi = \Omega_{\rm H}$. 

Substituting Eq.~(\ref{urh}) into Eqs.~(\ref{utd}) and (\ref{uphid}), 
we obtain the specific energy and the specific angular momentum of fluid 
particles as they reach the horizon
\begin{eqnarray}
	E_{\rm H} = 
		- f_{\rm E} + \frac{2M}{r_{\rm H}}\, c_0^2\, 
		\Omega_\Psi\left(\Omega_{\rm H}- 
		\Omega_\Psi\right)\;, \hspace{1cm}
	L_{\rm H} = 
		- f_{\rm L} + \frac{2M}{r_{\rm H}}\, c_0^2 
		\left(\Omega_{\rm H}- \Omega_\Psi\right)\;.	
		\label{lh} 	
\end{eqnarray} 

From Eq.~(\ref{urh}), $u_{\rm H}^r$ is always finite. However, $u_r u^r = 
(r^2/\Delta)(u^r)^2 \rightarrow \infty$ as $r\rightarrow r_{\rm H}$. Then, 
Eq.~(\ref{caa}) implies that $B^2/4\pi\rho_{\rm m} = c_{\rm A}^2/
\left(1-c_{\rm A}^2\right)$ is finite at $r = r_{\rm H}$. So we have
\begin{eqnarray}
        \left.\frac{c_{\rm A}^2/(1- c_{\rm A}^2)}{u_r u^r}
                \right\vert_{r \rightarrow r_{\rm H}} = 0 \;.
        \label{crurh2}
\end{eqnarray}
Equation~(\ref{crurh2}) implies that fluid particles must supersonically fall 
into the black hole.

The ratio of the electromagnetic energy density to the kinetic energy density,
as measured by a LNRF observer, can be calculated by
\begin{eqnarray}
     \zeta = \frac{B^\prime_a B^{\prime a} + E^\prime_a E^{\prime a}}{ 
	     8\pi\rho_{\rm m}\Gamma(\Gamma -1)} = \frac{1}{\Gamma(\Gamma 
	     -1)}\left[\frac{c_{\rm A}^2}{2\left(1- c_{\rm A}^2\right)} -
	     \frac{u^r}{\chi^2}\, c_0^2\left(\omega - \Omega_\Psi\right)^2
		\right] \;,
	\label{zeta}
\end{eqnarray}
where $B^{\prime a}$ and $E^{\prime a}$ are given by Eqs.~(\ref{mfield})
and (\ref{efield}), $\Gamma = - u_a U^b$ is the Lorentz factor of fluid 
particles. As $r\rightarrow r_{\rm H}$, we have $\Gamma\rightarrow\infty$ and
\begin{eqnarray}
     \zeta \rightarrow \zeta_{\rm H} \equiv
	     \frac{c_0^2 \left(-u^r_{\rm H}\right) \left(
	     \Omega_{\rm H} - \Omega_\Psi\right)^2}{\left(f_{\rm E} -
	     \Omega_{\rm H} f_{\rm L}\right)^2} \;,
	\label{zetah}
\end{eqnarray}
where we have used $\left(\Gamma\chi\right)_{r\rightarrow r_{\rm H}} = E_{\rm H} 
- L_{\rm H} \Omega_{\rm H}= -f_{\rm E} + \Omega_{\rm H} f_{\rm E}$\,. Hence, 
the ratio of the electromagnetic energy density to the kinetic energy density 
of the fluid is finite on the horizon of the black hole.

\subsection{Solutions at infinity}
\label{sec4.3}

In Sec.~\ref{sec3.1} we have shown that when $\Psi\neq 0$ the solutions have
a very bad dynamical behavior as $r\rightarrow\infty$: $B^2/4\pi\rho_{\rm
m} \propto (-u^r)^{-1}\rightarrow\infty$, indicating that solutions with 
nonzero $\Psi$ cannot extend to infinity. Therefore, here we set $\Psi = 
0$ to discuss the asymptotic behavior of the solutions at infinity. Then,
from Eq.~(\ref{fe}) we have $f_{\rm E} = - E$. We assume that $0<E\le 1$, then 
we have $f_{\rm E}^2 \le 1$.

When $f_{\rm E}^2 < 1$ (i.e., $0<E<1$), from Eq.~(\ref{funcf}) we have
$F(r,u^r) \approx \left(1 + c_0^2/r^2 u^r\right)^2\left(1 - f_{\rm E}^2\right) 
+ f_{\rm L}^2/r^2$\,, as $r\rightarrow\infty$. Since the right-hand side is 
always positive, solutions to $F = 0$ do not exist. This means that a flow with 
$f_{\rm E}^2 < 1$ cannot extend to infinity so must have an outer boundary of 
finite radius.

When $f_{\rm E}^2 = 1$ (i.e., $E = 1$), from Eq.~(\ref{funcf}) we have
\begin{eqnarray}
	F(r,u^r) \approx \left(1 + \frac{c_0^2}{r^2 u^r}\right)^2
		\left[(u^r)^2 - \frac{2M}{r}\right] + \frac{f_{\rm L}^2}
		{r^2} \;,
	\label{finf3}
\end{eqnarray}
as $r\rightarrow\infty$. Then, there are two possible solutions to $F = 0$ at 
infinity. One is $u^r \approx - (2M/r)^{1/2} + {\cal O}(r^{-3/2})$, which by 
Eq.~(\ref{uphiu}) leads to $r u^\phi\propto r^{-1}$ and
$|r u^\phi/u^r|\propto (M/r)^{1/2} \ll 1$. For the problem that here we are 
interested in, we expect that $|r u^\phi/u^r| \gg 1$ at large radii, so 
this asymptotic solution for $u^r$ is not what we are looking for.

The other asymptotic solution for $u^r$ is
\begin{eqnarray}
	u^r \approx - \frac{c_0^2} {r^2} + {\cal O}\left(\frac{1}{r^{5/2}}
		\right)\;, \hspace{1cm} r \rightarrow\infty \;.
	\label{urasp}
\end{eqnarray}
Then, from Eq.~(\ref{uphiu}), we have $r u^\phi \approx -(f_{\rm L}/r)/\left(
1+c_0^2/r^2 u^r\right) \approx \left[2M/r - (u^r)^2\right]^{1/2}$ as $r
\rightarrow\infty$,
where in the last step Eq.~(\ref{finf3}) and $F = 0$ have been used.
Then, the asymptotic $r u^\phi$ corresponding to Eq.~(\ref{urasp}) is
\begin{eqnarray}
	r u^\phi \approx \left(\frac{2M}{r}\right)^{1/2} \;,
	     \hspace{1cm} r \rightarrow\infty \;,
	\label{uphiasp}
\end{eqnarray}
which leads to the correct asymptotic behavior $\left|r u^\phi/u^r\right| 
\propto (r/M)^{3/2}\gg 1$. Note, Eq.~(\ref{uphiasp}) corresponds to a
super-Keplerian disk at large radii [a Keplerian disk has $r u^\phi = 
(M/r)^{1/2}$].

By Eq.~(\ref{rhm}), Eq.~(\ref{urasp}) implies that 
\begin{eqnarray}
	\rho_{\rm m}\approx \frac{1}{4\pi}\left(\frac{F_{\rm m}}{C_0}
		\right)^2= \mbox{const} \;,
	\hspace{1cm}r\rightarrow\infty \;.
	\label{rhoi}
\end{eqnarray}
Since $\Gamma \approx 1$, $\Gamma-1 \approx (r u^\phi)^2/2$, $B^{\prime 2} + 
E^{\prime 2} \approx B^2 \approx\left(C_0 u^\phi/r u^r\right)^2$ as $r
\rightarrow \infty$, from Eqs.~(\ref{urasp}), (\ref{rhoi}), and
(\ref{zeta}) we have 
\begin{eqnarray}
	\zeta \approx 1 \;, \hspace{1cm} \mbox{as $r\rightarrow\infty$} \;.
	\label{equi}
\end{eqnarray}
Thus, at large radii the energy of the magnetic field is about equipartitioned 
with the kinetic energy of the flow. This explains the super-Keplerian nature 
of the solutions: at large radii the magnetic field is dynamically important, 
the ``hoop stress'' of the toroidal magnetic field provides an inward force 
confining the flow in addition to the gravitational force.

The asymptotic solutions as $r\rightarrow r_{\rm H}$ and $\infty$, for the case 
of $\Psi = 0$ and $E = 1$, are summarized in Table~\ref{tab1}.

\section{Solutions with Constant Specific Energy}
\label{sec5}

From the analysis in Secs.~\ref{sec2} and \ref{sec3} we see that, for the 
model studied here, the integral constant $\Psi$ has to be zero 
in order that the solutions can extend to infinity with finite $r^2 B^r$ 
and $B^2/4\pi \rho_{\rm m}\,$. For a flow with $\Psi = 0$ (i.e., $\Omega_\Psi 
= 0$), the specific energy of fluid particles keeps constant, from $r = 
\infty$ down to $r = r_{\rm H}$. This type of solutions are of particular 
interest since they represent a new accretion mode in which accreting 
material loses its angular momentum but has its energy unchanged. 

From Sec.~\ref{sec4}, at large radii the flow must be 
in the subsonic state since $u_r u^r (1-c_{\rm A}^2)/ c_{\rm A}^2 \sim 
(M/r)^3\ll 1\,$. On the other
hand, the flow must be in a supersonic state as it reaches the horizon of
the black hole [Eq.~(\ref{crurh2})]. Therefore, a fast critical point 
[defined by Eq.~(\ref{rf})] must exist somewhere between $r= r_{\rm H}$
and $r = \infty$. Like in all wind problems, any physical solution must 
pass the fast critical point smoothly. 

In this section we solve the radial momentum equation to look for solutions 
corresponding to a flow accreting from infinity onto a black hole with 
constant specific energy.

\subsection{Solutions around a Schwarzschild black hole}
\label{sec5.1}

We study the simplest case first: solutions with constant specific 
energy around a Schwarzschild black hole (i.e., $a = 0$). The effects of the 
black hole spin on the solutions are studied in the next subsection.

When $a= \Psi = 0$, the generation function [Eq.~(\ref{funcf})] becomes
\begin{eqnarray}
        F\left(r,u^r\right)\equiv
                \left[1+ \frac{c_0^2}{r^2 u^r}\left(1 -\frac{2M}{r}
                \right)\right]^2\,\left[\frac{\left(u^r\right)^2-
		f_{\rm E}^2}{1 -\frac{2M}{r}} + 1\right] + 
		\frac{f_{\rm L}^2}{r^2} \;.
        \label{funcf2}
\end{eqnarray}
The corresponding derivatives are
\begin{eqnarray}
	\frac{\partial F}{\partial u^r} &=& \frac{2}{u^r}
		\left[1+ \frac{c_0^2}{r^2 u^r}\left(1 -\frac{2M}{r}
                \right)\right]\left[\frac{\left(u^r\right)^2}{1 -
		\frac{2M}{r}} -\frac{c_0^2}{r^2 u^r}\left(1 -
		\frac{2M}{r}-f_{\rm E}^2\right)\right] \;,
		\label{dfur}\\
	\frac{\partial F}{\partial r} &=& \frac{2}{r}
		\left[1+ \frac{c_0^2}{r^2 u^r}\left(1 -\frac{2M}{r}
                \right)\right]\left\{-\frac{\left(u^r\right)^2
		-f_{\rm E}^2}{\left(1 -\frac{2M}{r}\right)^2} 
		\left[\frac{M}{r}+\frac{c_0^2}{r^2 u^r}\left(1 -
		\frac{2M}{r}\right)\left(2-\frac{5M}{r}\right)
		\right]\right.
		\nonumber\\
		&&\left. -\frac{2 c_0^2}{r^2 u^r}\left(1-\frac{3M}{r}
		\right)\right\} - \frac{2 f_{\rm L}^2}{r^3}
		\label{dfr}\;.
\end{eqnarray}

As shown in Sec.~\ref{sec4.1}, the only singular point in the radial momentum 
equation is the fast critical point that is defined by
\begin{eqnarray}
	\frac{\left(u^r\right)^2}{1 -
		\frac{2M}{r}} -\frac{c_0^2}{r^2 u^r}\left(1 -
		\frac{2M}{r}-f_{\rm E}^2\right) = 0 \;,
		\hspace{1cm}
		r = r_{\rm f} \;,
	\label{rf2}
\end{eqnarray}
from which we can solve for the radial velocity at $r =  r_{\rm f}$
\begin{eqnarray}
	u_{\rm f}^r = -\left[\frac{c_0^2}{r_{\rm f}^2}\left(1 -
		\frac{2M}{r_{\rm f}}\right)\left(-1 +\frac{2M}
		{r_{\rm f}}+f_{\rm E}^2\right)\right]^{1/3} \;.
	\label{vf2}
\end{eqnarray}
Since $u_{\rm f}^r$ must be negative and $r_{\rm f} > r_{\rm H} = 2M$, the 
following condition must be satisfied
\begin{eqnarray}
	-1 + \frac{2M}{r_{\rm f}} + f_{\rm E}^2 > 0 \;.
	\label{url0}
\end{eqnarray}
Note that $u_{\rm f}^r$ does not depend on $f_{\rm L}$.

From the integral radial momentum equation $F(r,u^r) = 0$, we can solve for 
$f_{\rm L}$
\begin{eqnarray}
        f_{\rm L} = - r
                \left[1+ \frac{c_0^2}{r^2 u^r}\left(1 -\frac{2M}{r}
                \right)\right]\,\left[-\frac{\left(u^r\right)^2-
		f_{\rm E}^2}{1 -\frac{2M}{r}} - 1\right]^{1/2} \;.
        \label{solfl}
\end{eqnarray}
Substituting Eq.~(\ref{solfl}) into Eq.~(\ref{dfr}), we can eliminate $f_{\rm L}$ 
from Eq.~(\ref{dfr}) and obtain
\begin{eqnarray}
	\frac{\partial F}{\partial r} &=& \frac{2}{r}
		\left[1+ \frac{c_0^2}{r^2 u^r}\left(1 -\frac{2M}{r}
                \right)\right]\left\{1 -\frac{c_0^2}{r^2 u^r}
		\left(1-\frac{4M}{r}\right)\right. \nonumber\\
		&&\left.+\frac{\left(u^r\right)^2
		-f_{\rm E}^2}{\left(1 -\frac{2M}{r}\right)^2} 
		\left(1-\frac{3M}{r}\right)\left[1-\frac{c_0^2}{r^2 
		u^r}\left(1 -\frac{2M}{r}\right)\right]\right\}
		\label{dfr2}\;.
\end{eqnarray}

For $f_{\rm L}$ to be real, the term in the second pair of brackets on the 
right-hand side of Eq.~(\ref{solfl}) must be nonnegative
\begin{eqnarray}
	-\frac{\left(u^r\right)^2-
		f_{\rm E}^2}{1 -\frac{2M}{r}} - 1 \ge 0 \;.
	\label{con1a}
\end{eqnarray}
Making use of Eqs.~(\ref{vf2}) and (\ref{url0}), at $r = r_{\rm f}$ 
Eq.~(\ref{con1a}) becomes
\begin{eqnarray}
	\left(1 - \frac{2M}{r_{\rm f}}\right)^2\, \frac{c_0^4}
		{r_{\rm f}^4} \le -1 + \frac{2M}{r_{\rm f}} + 
		f_{\rm E}^2 \;.
	\label{con1}
\end{eqnarray}
Equation~(\ref{con1}) restricts the region in the $(r_{\rm f},c_0)$ 
space where physical solutions exist.

At the fast critical point Eq.~(\ref{rcuc}) must be satisfied. From 
Eq.~(\ref{dfur}), at $r = r_{\rm f}\,$ $\partial F/\partial u^r = 0$ 
leads to Eq.~(\ref{vf2}). From Eq.~(\ref{dfr2}), at $r = r_{\rm 
f}\,$ $\partial F/\partial r = 0$ is equivalent to
\begin{eqnarray}
	1 -\frac{c_0^2}{r_{\rm f}^2 u_{\rm f}^r}\left(1-\frac{4M}
	        {r_{\rm f}}\right)+\frac{\left(u_{\rm f}^r\right)^2
		-f_{\rm E}^2}{\left(1 -\frac{2M}{r_{\rm f}}\right)^2} 
		\left(1-\frac{3M}{r_{\rm f}}\right)\left[1 -\frac{c_0^2}
		{r_{\rm f}^2 u_{\rm f}^r}\left(1 -\frac{2M}{r_{\rm f}}
		\right)\right]= 0
		\label{dfr3}\;.
\end{eqnarray}

From Eqs.~(\ref{vf2}) and (\ref{dfr3}) we can numerically solve for $r_{\rm f} 
= r_{\rm f}\left(f_{\rm E}, c_0\right)$ and $u_{\rm f}^r = u_{\rm f}^r\left(
f_{\rm E}, c_0\right)$ for any given $f_{\rm E}$ and $c_0$. The location of the 
fast critical point in the $(r,u^r)$ phase space is then determined. 
Substituting the solutions into Eq.~(\ref{solfl}), we can determine $f_{\rm L} 
= f_{\rm L}\left(f_{\rm E}, c_0\right)$, which has a negative sign.

From Eq.~(\ref{fe}), we have $f_{\rm E} = -E$ when $\Psi = 0$. When the flow 
has an outer boundary with a finite radius $r_0$, we assume that the constant 
specific energy of fluid particles, $E$, is equal to that of a test 
particle moving on a Keplerian circular orbit at $r = r_0$. Then, we have
$f_{\rm E} = -E_{\rm Keplerian}(r_0)$. When $r_0 = \infty$, we assume that 
$f_{\rm E} = -E = -1$.

\begin{figure}
\includegraphics[width=15.2cm]{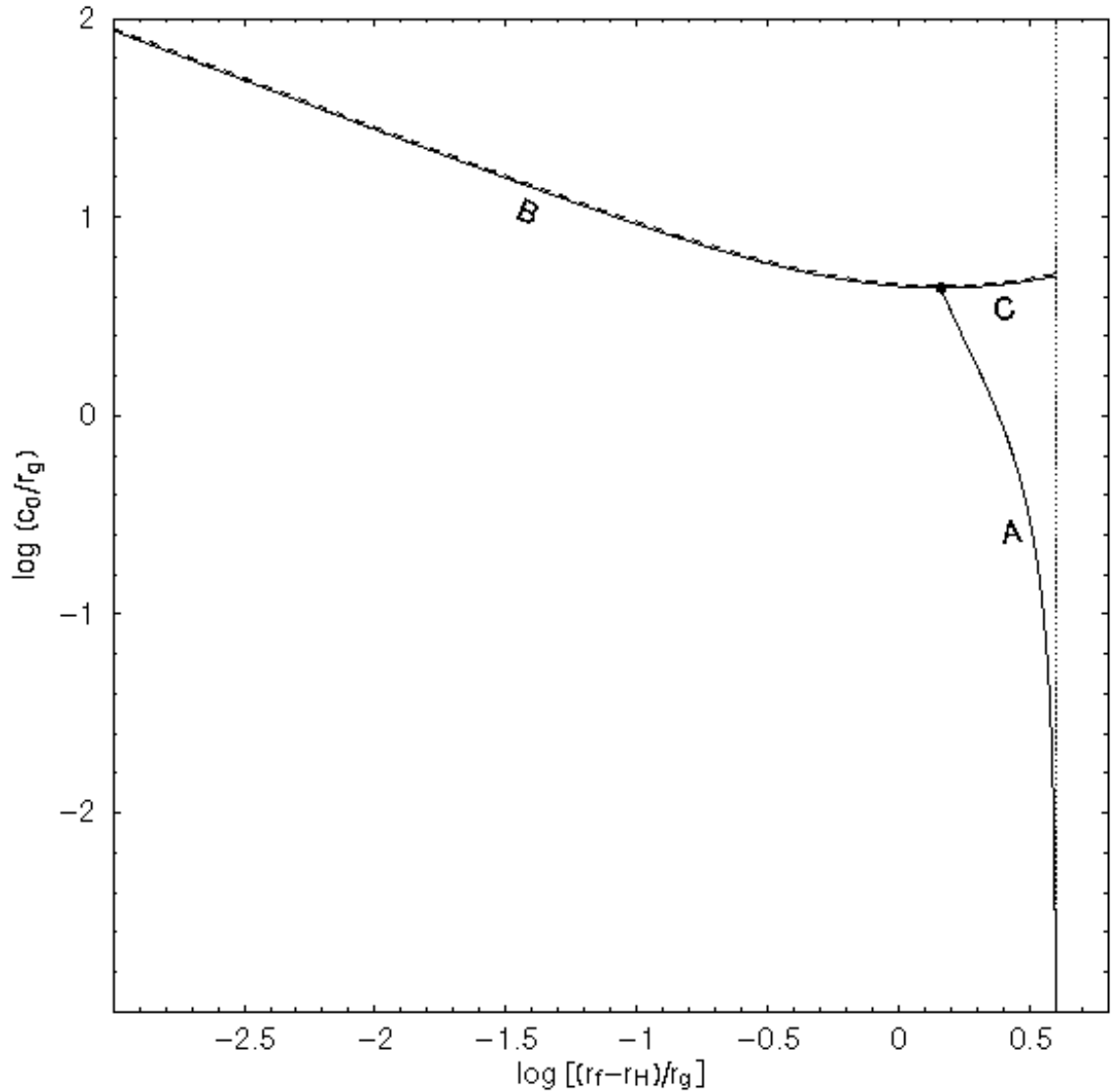}
\caption{Solutions for $r_{\rm f}$---the radius at the fast critical point, 
corresponding to a magnetized flow with constant specific energy accreting 
onto a Schwarzschild black hole (solid curves). The parameter $c_0$, 
defined by Eq.~(\ref{c0etc}), is proportional to the flux of radial 
magnetic field. The outer boundary of the flow is at the marginally stable 
circular orbit: $r_0 = r_{\rm ms} = 6\, r_{\rm g}$ (dotted line), where 
$r_{\rm g} \equiv M$ is the gravitational radius. The specific energy of 
fluid particles is equal to that of a test particle moving on the 
marginally stable circular orbit. The solutions for $r_{\rm f}$ consist of 
three branches: A, B, and C, with intersection 
at the dark point. The dashed curve represents the boundary for physical 
solutions: beyond the dashed curve physical solutions do not exist (see 
the text). (The dashed curve is indeed coincident with the 
solid curves B and C. To make it visible, we have shifted the dashed curve
upward a little bit.)  
\label{fig1}}
\end{figure}

In Fig.~\ref{fig1} we show the solutions for $r_{\rm f}$ corresponding to 
$r_0 = r_{\rm ms}$\,, where $r_{\rm ms}= 6\, r_{\rm g}$ is the radius at the
marginally stable circular orbit \cite{bar72}, $r_{\rm g}\equiv M$ is the 
gravitational radius of the black hole. This represents the solutions in the 
plunging region from the inner boundary of a thin Keplerian disk to the 
horizon of the black hole. The solutions for $r_{\rm f}$, represented by 
the solid curves in the figure, consist of three branches: A, B, and C, 
intersecting 
at the dark point. The dashed curve represents the boundary for physical 
solutions to exist: above the dashed curve Eq.~(\ref{con1}) is violated so 
physical solutions do not exist. (The dashed curve is indeed coincident with 
the solid curves B and C. To make it visible, in the figure we have shifted 
the dashed curve upward a little bit.) From Fig.~\ref{fig1} we see that, below 
the dark point (where $r = 3.4477\,r_{\rm g}$, $c_0 = 4.4030\,r_{\rm g}$), 
$r_{\rm f}$ is uniquely determined by the value of $c_0$. As $c_0^2/r_{\rm 
g}^2 \rightarrow 0\,$, $r_{\rm f}$ approaches the radius at the marginally 
stable circular orbit $r_{\rm ms}$ (the vertical dotted line). This 
is just what we expect for a weakly magnetized thin Keplerian disk: the 
(magneto)sonic point should not be far from the marginally stable circular 
orbit (\cite{muc82,abr89,afs02} and references therein). Above the dark point, 
multiple solutions for $r_{\rm f}$ exist. In this paper we focus on the 
situation of weak magnetic fields with $c_0^2/r_{\rm g}^2 \ll 1$, 
so we do not discuss the solutions above the dark point in detail.

\begin{figure}
\vspace{2cm}
\includegraphics[width=15.2cm]{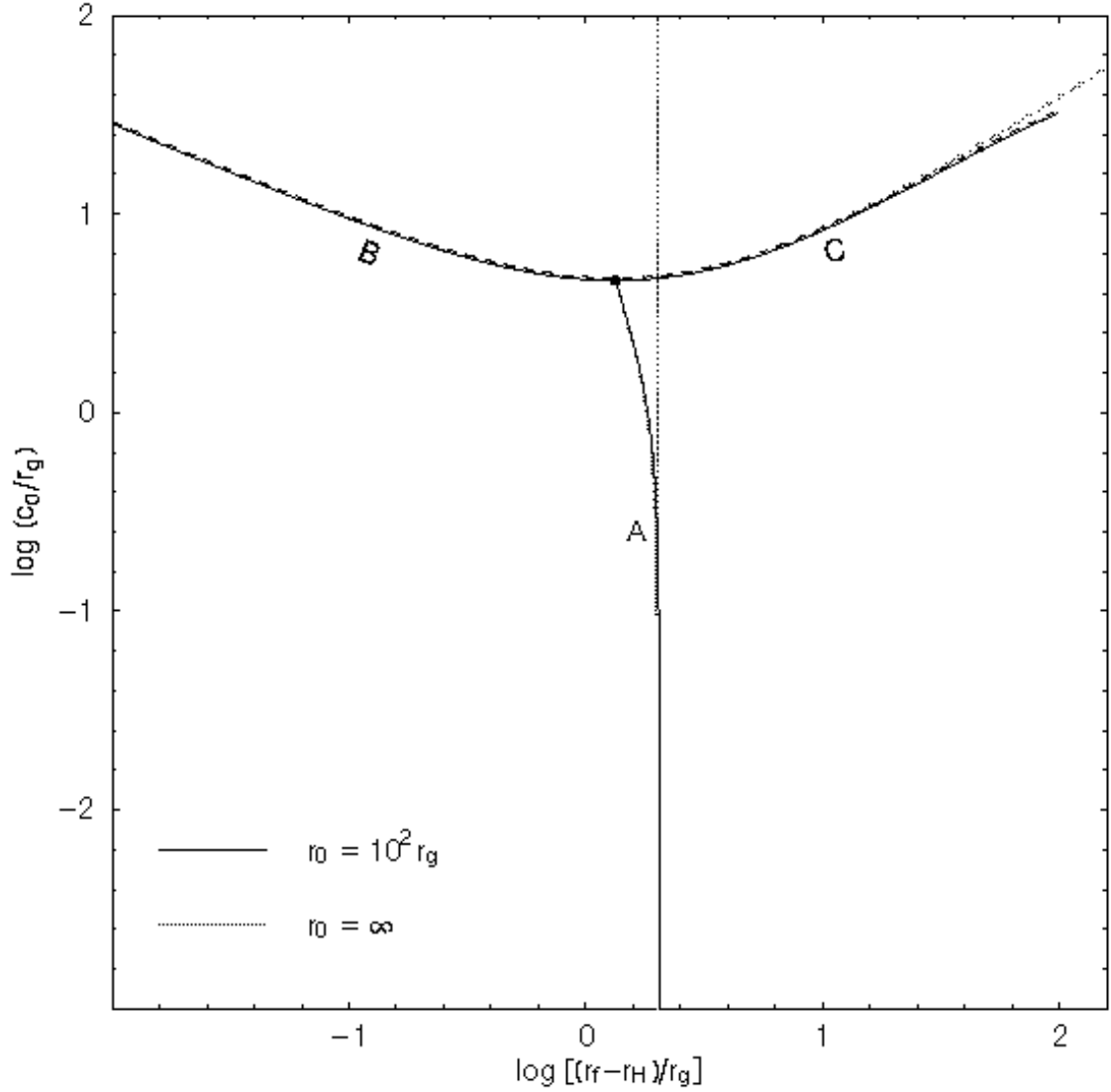}
\caption{Same as Fig.~\ref{fig1} but for $r_0 = 10^2 r_{\rm g}$ (dark lines)
and $r_0 = \infty$ (gray lines). The specific energy of fluid particles
is equal to that of a test particle moving on a Keplerian circular orbit 
at $r = r_0$. The vertical dotted line marks the location of the marginally 
bound circular orbit ($r_{\rm mb} = 4\, r_{\rm g}$).
\label{fig2}}
\end{figure}

We are more interested in solutions with $r_0 \gg r_{\rm g}$, corresponding 
to a magnetized flow accreting onto a black hole from large radii. In
Fig.~\ref{fig2} we show the solutions for $r_{\rm f}$ corresponding to 
$r_0 = 10^2 r_{\rm g}$ (dark lines) and $r_0 = \infty$ (gray lines). The 
solutions for $r_0 \gg r_{\rm g}$ have the same topology as that for $r_0
= r_{\rm ms}$. However, when $c_0/r_{\rm g}$ is fixed, $r_{\rm f}$ decreases 
as $r_0$ increases. For $r_0 \gg r_{\rm g}$, $r_{\rm f}$ becomes insensitive 
to the value of $r_0$: on branches A and B the solutions for $r_0 = 10^2 
r_{\rm g}$ are almost indistinguishable from those for $r_0 = \infty$. This 
is caused by the fact that $f_{\rm E} \approx -1$ 
for $r_0 \gg r_{\rm g}$. From Fig.~\ref{fig2} we see that, for the case of 
$r_0 \gg r_{\rm g}$, $r_{\rm f}$ approaches the radius at the marginally bound 
circular orbit $r_{\rm mb} = 4\,r_{\rm g}$ \cite{bar72} (the 
vertical dotted line) in the limit $c_0^2/r_{\rm g}^2\rightarrow 0\,$.

In both Figs.~\ref{fig1} and \ref{fig2} the dashed curve is coincident 
with the solution branches B and C. So, the solution branches B and C
correspond to $f_{\rm L} = 0$ [see Eqs.~(\ref{solfl}) and (\ref{con1a})]. 
By Eq.~(\ref{lh}), such solutions correspond to $L_{\rm H} =
0$ since $\Omega_\Psi = \Omega_{\rm H} = 0$. Thus, for a sufficiently strong 
magnetic field (corresponding to the branches B and C), fluid particles lose 
all of their angular momentum as they reach the horizon of the black hole.

Once $r_{\rm f}$, $u_{\rm f}^r$, and $f_{\rm L}$ are determined as functions 
of $c_0$ and $f_{\rm E}$, we can solve the radial momentum equation for the 
radial velocity $u^r$ as a function of radius 
$r$. However, generally it is not a simple thing to solve the algebraic 
equation $F(r,u^r) =0$, which has singularities and multiple roots. In 
practice we would rather solve the first-order differential equation
\begin{eqnarray}
	\frac{d\ln u^r}{d\ln r} = - \frac{1 -\frac{c_0^2}{r^2 u^r}
		\left(1-\frac{4M}{r}\right)+\left[\left(u^r\right)^2
		-f_{\rm E}^2\right]\left(1 -\frac{2M}{r}\right)^{-2} 
		\left(1-\frac{3M}{r}\right)\left[1 -\frac{c_0^2}
		{r^2 u^r}\left(1 -\frac{2M}{r}\right)\right]}
		{\left(u^r\right)^2 \left(1 - \frac{2M}{r}\right)^{-1}
		-\frac{c_0^2}{r^2 u^r}\left(1 -
		\frac{2M}{r}-f_{\rm E}^2\right)} \;, \nonumber\\
	\label{durr}
\end{eqnarray}
which is obtained by substituting Eqs.~(\ref{dfur}) and (\ref{dfr2})
into Eq.~(\ref{deq}). Note that the only singularity in Eq.~(\ref{durr}) 
is the fast critical point.

We choose on each side of the fast critical point a pair of $(r_i,u_i^r)$,
which can be obtained by solving the algebraic equation $F(r, u^r) =0$ 
where initial guessed values can be obtained by visually checking the
contour plot of $F$ versus $r$ and $u^r$. Then, starting from the point
$(r_i,u_i^r)$, we integrate Eq.~(\ref{durr}) inward and outward, toward the 
horizon of the black hole and the fast critical point if $(r_i,u_i^r)$ is 
on the left-hand side of the fast critical point, or toward the fast 
critical point and large radii if $(r_i,u_i^r)$ is on the right-hand side 
of the fast critical point. [The starting point $(r_i,u_i^r)$cannot be exactly 
at the fast critical point since the latter is a fixed point of the 
differential equation.] 

\begin{figure}
\vspace{2cm}
\includegraphics[width=15.6cm]{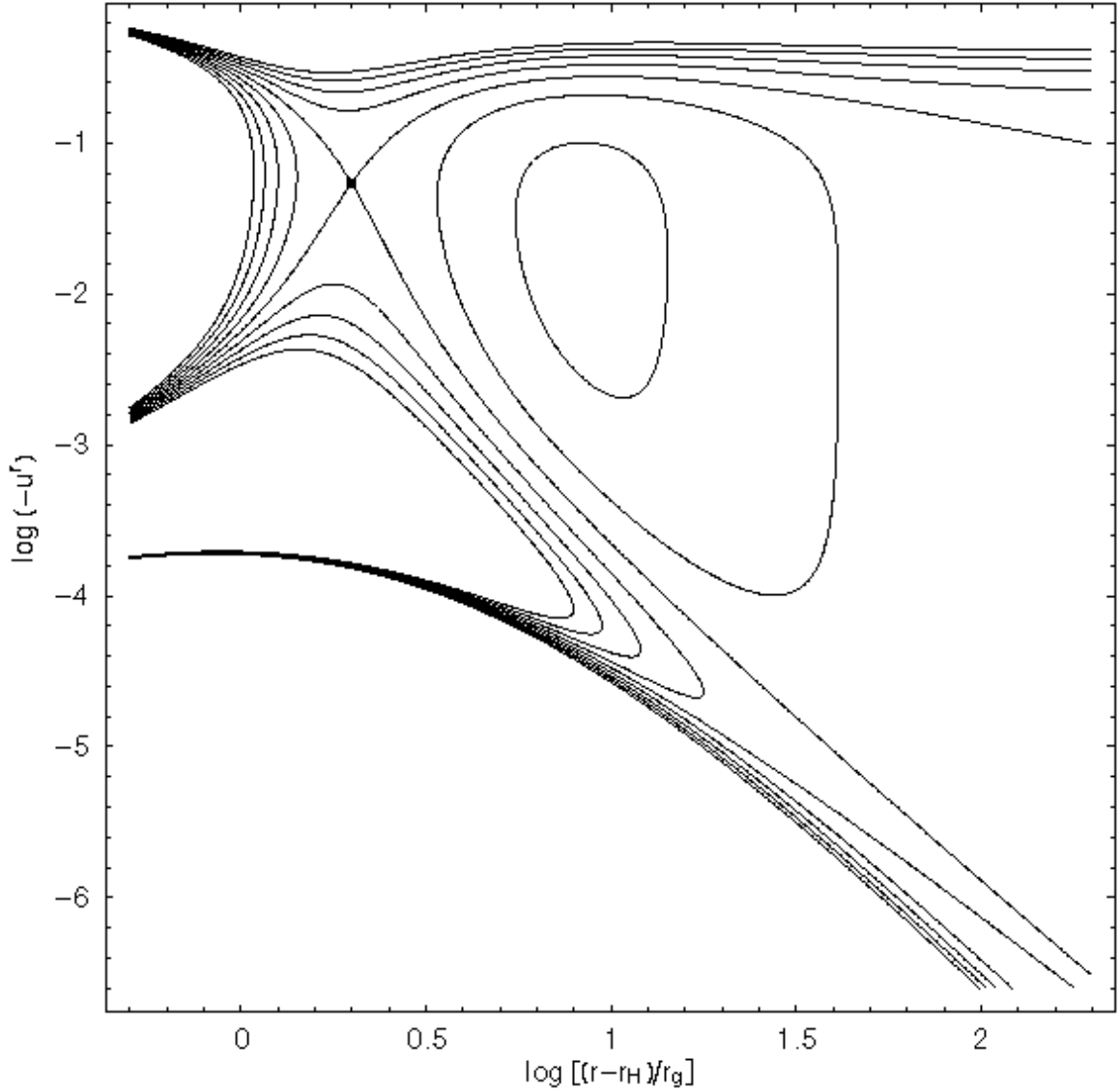}
\caption{Contour plots of the generation function~(\ref{funcf2}) versus 
$r$ and $u^r$. The parameters are: $c_0 = 0.1\, r_{\rm g}$, $f_{\rm E} = 
- 1$, $f_{\rm L}= -3.96522\, r_{\rm g}$ (corresponding to $r_0 = \infty$). 
Then, the fast critical point, which corresponds to a saddle point of the 
generation function, is at $(r_{\rm f}, u_{\rm f}^r) = (3.98852\, r_{\rm 
g}, -0.053964)$, as indicated by the dark point. Contours are plotted from 
$F = -0.2$ to $F = 0.2$ with step $\delta F = 0.05$. The contours of $F = 
0$ are given by the two lines that smoothly pass the fast critical point,
the one that goes from the right-bottom corner to the left-top corner
represents the physical solution for the accretion flow.
\label{fig3}}
\end{figure}

As an example, we have solved the equation with $f_{\rm E} = -1$ (i.e., $r_0 
= \infty$) and $c_0 = 0.1\, r_{\rm g}$, corresponding to $r_{\rm f} = 3.98852\, 
r_{\rm g}$\,, $u_{\rm f}^r = -0.053964$\,, and $f_{\rm L} =-3.96522\, 
r_{\rm g}$\,. The contours of the corresponding generation function are plotted
in Fig.~\ref{fig3}~\footnote{The contour plot is drawn with {\it 
Mathematica}.}, from $F = -0.2$ to $F = 0.2$ with step $\delta F = 0.05$. The 
contours of $F = 0$, which correspond to the
physical solutions, are the two diagonal lines that smoothly pass the 
fast critical point at the dark point. The solution that we are 
interested in here should have $0<-u^r\ll 1$ at large radii and $u^r \sim 
-1$ near $r = r_{\rm H}$, which corresponds to the diagonal line that goes 
from the right-bottom corner to the left-top corner in Fig.~\ref{fig3}. 
The differential radial momentum equation~(\ref{durr}) is then solved along 
this diagonal line, with the approach described above.

\begin{figure}
\includegraphics[width=14.5cm]{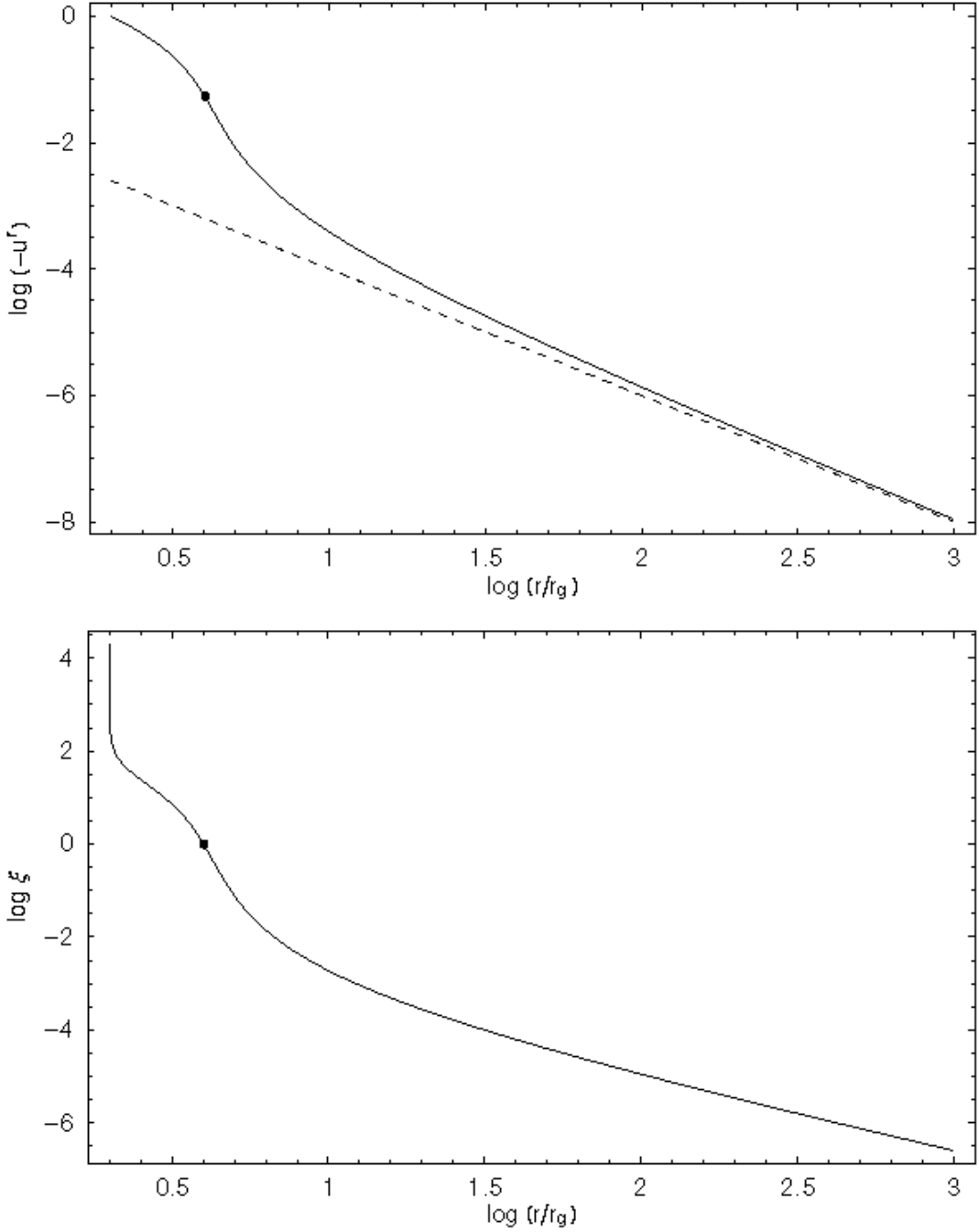}
\caption{Upper panel: 
Solution to the radial momentum equation~(\ref{durr}). The parameters
are: $c_0 = 0.1\, r_{\rm g}$, $f_{\rm E} = - 1$, and $f_{\rm L}= -3.96522\,
r_{\rm g}$ (corresponding to $r_0 = \infty$). The dark point shows the location
of the fast critical point. The dashed line shows the asymptotic solution $u^r 
= - c_0^2/r^2$. Lower panel: The ratio $\xi = \left[u_r u^r\left(1-c_{\rm 
A}^2\right)/c_{\rm A}^2\right]^{1/2}$ corresponding to the solution in 
the upper panel. Subsonic motion
corresponds to $\xi < 1$. Supersonic motion corresponds to $\xi>1$. The fast 
critical point is at $\xi =1$, as indicated by the dark point.
\label{fig4}}
\end{figure}

The solution for the radial velocity $u^r$ as a function of $r$ is plotted 
in Fig.~\ref{fig4} (upper panel). For comparison, we show the asymptotic 
solution as $r\rightarrow\infty$ with the dashed line [see Eq.~(\ref{urasp})]. 
We have calculated $\xi \equiv \left[u_r u^r\left(1-c_{\rm A}^2\right)/
c_{\rm A}^2\right]^{1/2}$, and shown the results in the lower panel of 
Fig.~\ref{fig4}. The parameter $\xi$ measures the ratio of the radial velocity 
to the Alfv\'{e}n velocity. At the fast critical point we have $\xi =1$ 
[Eq.~(\ref{rf})]. The 
figure clearly shows that the flow starts from a subsonic state ($\xi <1$)
at large $r$, passes the fast critical point at $r = r_{\rm f}$ ($\xi = 1$), 
then enters a supersonic state ($\xi > 1$) at $r< r_{\rm f}$. The ratio 
$\xi$ blows up at $r = r_{\rm H}$, confirming our analytic analysis in 
Sec.~\ref{sec4.2} [see Eq.~(\ref{crurh2})].

\begin{figure}
\vspace{2cm}
\includegraphics[width=15.5cm]{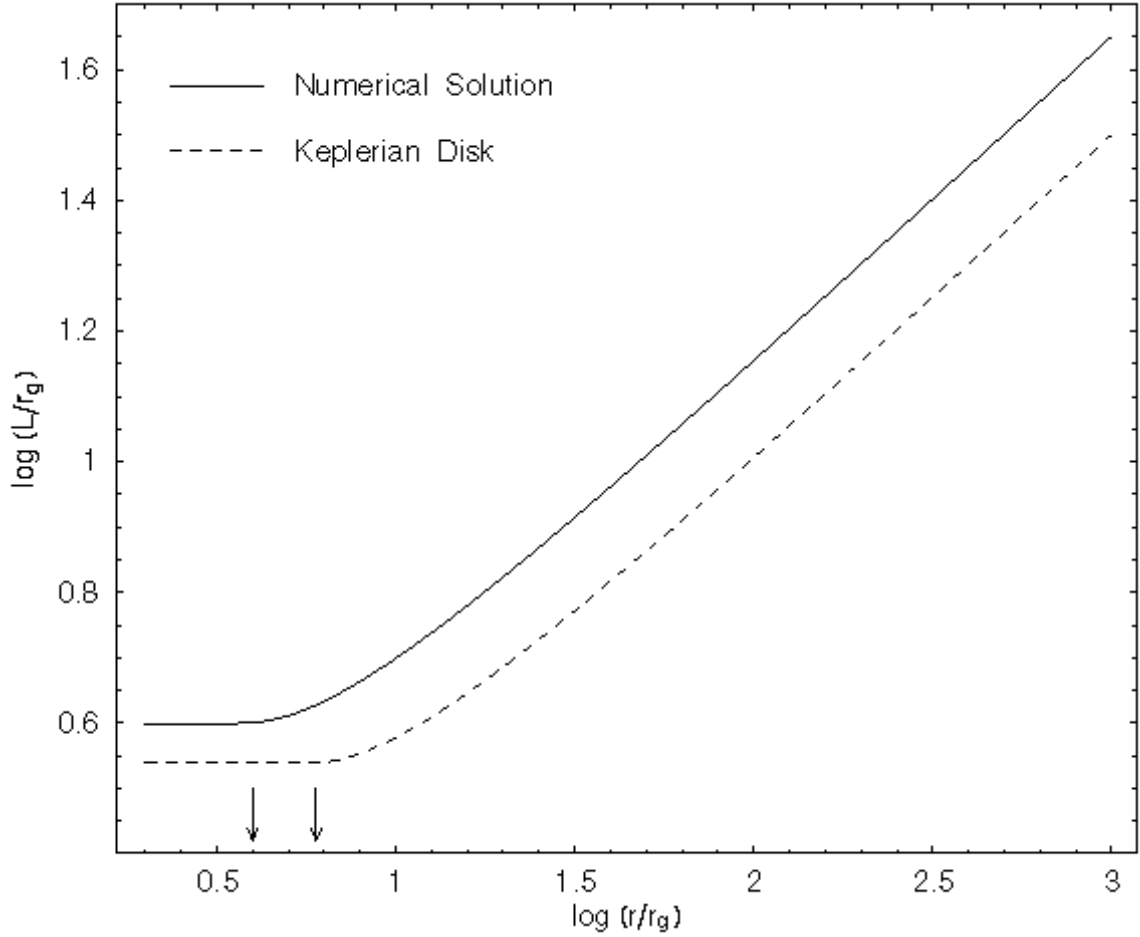}
\caption{Evolution of the specific angular momentum of the particles in 
the flow described
by the solution in Fig.~\ref{fig4} (solid line). The dashed line shows the 
solution corresponding to a thin Keplerian disk joined to a free-fall flow
inside the marginally stable circular orbit. The two arrows show the 
locations of the fast critical point (left arrow) corresponding to the
solution in Fig.~\ref{fig4}, and the marginally stable circular orbit 
(right arrow).  
\label{fig5}}
\end{figure}

The specific angular momentum of the fluid particles, $L$, can be 
calculated by substituting $a = \Psi = 0$, and the solution $u^r(r)$ 
into Eq.~(\ref{uphid}). The results are shown in Fig.~\ref{fig5}. For 
comparison, we show the specific angular momentum of a thin Keplerian disk 
with the dashed line. The Keplerian disk is joined to a free-fall flow 
in the plunging region, so its specific angular momentum keeps constant 
inside the marginally stable circular orbit. The fast critical point and the 
marginally stable circular
orbit are indicated by two arrows (left and right, respectively). 
From Fig.~\ref{fig5} we see that, a disk flow with constant specific energy 
can lose angular momentum at a rate similar to that of a thin Keplerian 
disk.

\begin{figure}
\vspace{2cm}
\includegraphics[width=15.5cm]{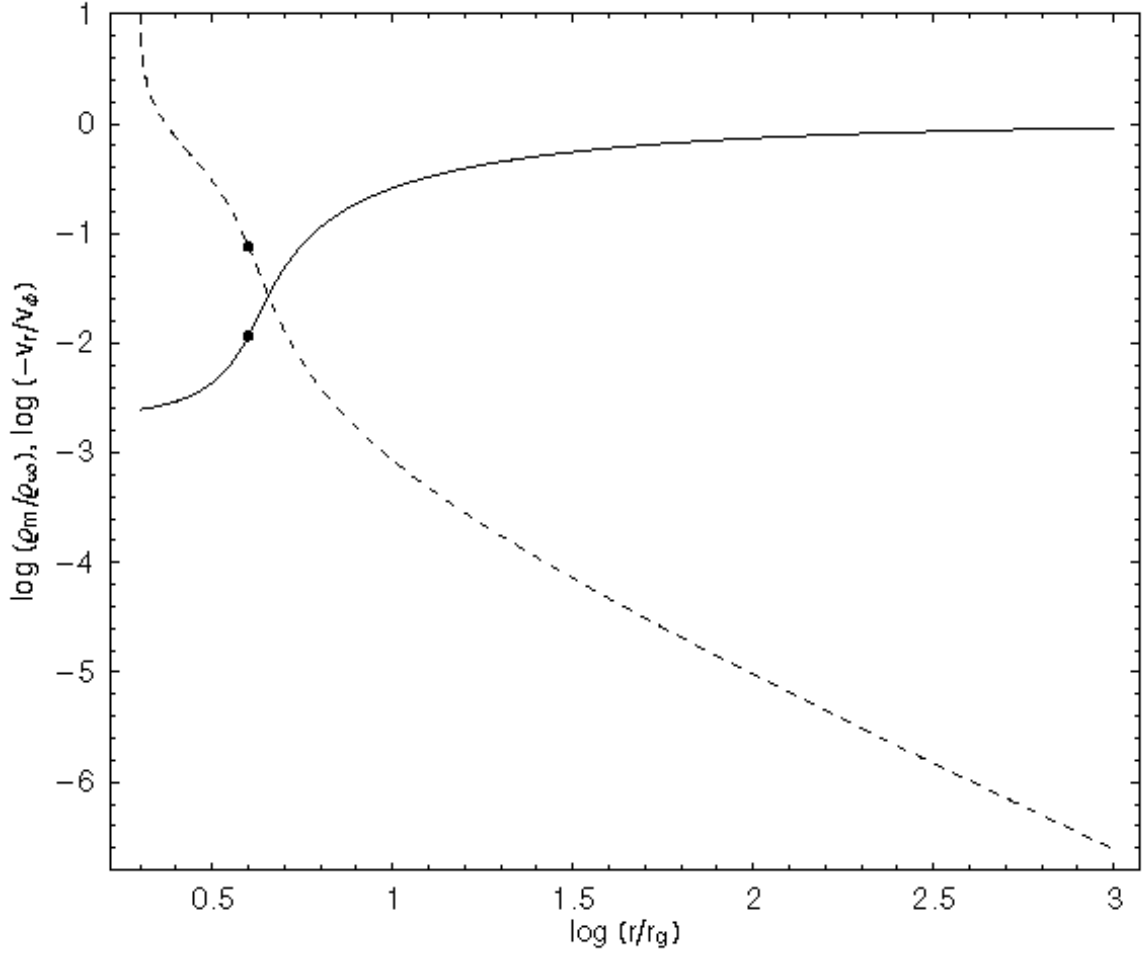}
\caption{Evolution of the mass density (solid line) and the ratio of the
radial velocity to the rotational velocity (dashed line), for the flow
described by the solution in Fig.~\ref{fig4}. The mass density is in units 
of its value at infinity: $\rho_\infty \equiv F_{\rm m}^2/\left(4\pi C_0^2
\right)$ [see Eq.~(\ref{rhoi})]. The dark points show the location of the 
fast critical point.
\label{fig6}}
\end{figure}

Figure~\ref{fig5} shows that the specific angular momentum of the 
flow remains almost constant after the flow passes the fast critical point. 
This suggests that we can treat the fast critical point as an inner boundary 
of the accretion flow. To justify this, we have calculated the mass 
density of the flow and the ratio of the radial velocity to the rotational 
velocity, as functions of radius $r$. The mass density of the flow is 
simply given by Eq.~(\ref{rhm}), which approaches a constant as $r
\rightarrow\infty$ [Eq.~(\ref{rhoi})]. The ratio of the radial velocity to 
the rotational velocity, as measured by a local static observer, is given 
by $v_r/v_\phi = r u^r/\chi L$\,. The numerical
results for the mass density and the ratio of the radial velocity to the 
rotational velocity, corresponding to the solution in Fig.~\ref{fig4}, are 
shown in Fig.~\ref{fig6}. 

We see that, at large radii $|v_r/v_\phi| \ll 1$, and 
$\rho_{\rm m}\rightarrow\rho_\infty\equiv F_{\rm m}^2/\left(4\pi C_0^2
\right)$ [Eq.~(\ref{rhoi})]. As the flow gets close to and passes the fast 
critical point (the dark points in the figure), $\rho_{\rm m}/\rho_\infty$ 
drops quickly and $|v_r/v_\phi|$ increases quickly. At the 
fast critical point we have $\rho_{\rm m} = 0.01167\, \rho_\infty$ and
$v_r = - 0.07629\, v_\phi$\,.
These numbers support the point that the fast critical point behaves 
like an inner boundary of the disk accretion flow: it marks the transition 
of the flow from a state with $|v_r/v_\phi| \ll 1$ at large 
radii to a state with $|v_r/v_\phi| \agt 1$ near 
the horizon of the black hole. Beyond the inner boundary, the magnetic field
plays an important role in the dynamics of the flow and transports angular
momentum outward. Inside the inner boundary, the dynamical effects of the
magnetic field become unimportant and the fluid particles free-fall to the
horizon of the black hole (Fig.~\ref{fig5}).

Figure~\ref{fig6} also shows that, as $r\rightarrow r_{\rm H}$, $\rho_{\rm 
m}$ is finite but $|v_r/v_\phi|$ blows up. The former is because of the fact 
that $u^r$ is finite at $r = r_{\rm H}$ (see Sec.~\ref{sec4.2}), the latter 
is caused by the well-known fact that $v_\phi\rightarrow 0$ and $v_r 
\rightarrow -1$ as particles approach the horizon.

The ratio of the magnetic energy density to the kinetic energy density of the 
flow, as measured by a local static observer [see Eq.~(\ref{zeta}); note that
when $a = \Psi =0$ the electric field is zero, see Eq.~(\ref{efield})], is
calculated and shown in Fig.~\ref{fig7}. In the region of $r\gg r_{\rm f}$, 
the ratio $\zeta$ is $\approx 1$, i.e., the energy of the magnetic
field is about equipartitioned with the kinetic energy of the flow 
[Eq.~(\ref{equi})]. This is caused by the fact that for $r\gg r_{\rm f}$ the 
velocity of the flow is predominantly rotational: $|v_r/v_\phi| 
\ll 1$ (Fig.~\ref{fig6}), so the magnetic field is sufficiently amplified 
by the shear rotation until the energy density of the magnetic field is 
comparable to the kinetic energy density of the flow. In the region of $r \alt 
r_{\rm f}$, we have $\zeta \ll 1$, caused by the fact that for $r \alt 
r_{\rm f}$ the radial component of the flow velocity becomes comparable to, 
or even greater than the azimuthal component (Fig.~\ref{fig6}), then the 
amplification of the magnetic field by the shear motion becomes 
unimportant \cite{li03}. At the fast critical point (the dark point in the 
figure), we have $\zeta = 0.004954\,$. As $r\rightarrow r_{\rm H}$, we have 
$\zeta\rightarrow 0\,$.

To see how sensitive the solutions are to the variation of $c_0$, and how the 
solutions approach the asymptotic solutions at infinity found in 
Sec.~\ref{sec4.3}, we have solved the radial momentum equation with $f_{\rm E} 
= - 1$ corresponding to two different values of $c_0$: $c_0 = 1\, r_{\rm g}$ 
and $c_0 = 10^{-2}r_{\rm g}$. The solutions for the angular velocity $\Omega = 
u^\phi/u^t$ are shown in Fig.~\ref{fig8}. The solid line corresponds to $c_0 = 
1\,r_{\rm g}$, the dashed line to $c_0 = 10^{-2}r_{\rm g}$. The 
dotted line in the figure is the angular velocity of a thin Keplerian disk 
joined to a free-fall flow in the plunging region. The dashed-dotted line 
shows the asymptotic solution for $\Omega$ derived from 
Eq.~(\ref{uphiasp}). We see that, 
the solutions are insensitive to the value of $c_0$ for $c_0^2 \alt r_{\rm 
g}^2$, and they are well represented by the asymptotic solution in the region 
with $\log(r/r_{\rm g}) > 1.5\,$. 

\subsection{Solutions around a Kerr black hole}
\label{sec5.2}

The spin of the black hole has important effects on the solutions of an 
accretion flow near the horizon of the black hole. For example, for the case 
of a thin Keplerian disk, the spin of the black hole determines the location 
of the inner boundary of the disk, how much energy is dissipated in the disk, 
and how much angular momentum is carried into the black hole by the flow 
\cite{nov73,pag74,tho74}. In this subsection we show how the spin of the 
black hole affects the accretion process by presenting some solutions for 
a magnetized flow with constant specific energy accreting onto a Kerr black 
hole. We assume that at large radii the specific angular momentum of the disk 
is always positive, but the specific spinning angular momentum of the black 
hole, $a$, can be either positive or negative.

For the case of a Kerr black hole with $\Psi = 0$, the generation function
becomes
\begin{eqnarray}
	F(r,u^r) &=& \left[1 + \frac{c_0^2}{r^2 u^r}\left(1-\frac{2M}
		{r}\right)\right]^2 \left[\frac{r^2}{\Delta} \left(u^r
		\right)^2 + 1 -\frac{f_{\rm E}^2} {1- \frac{2M}{r}}
		\right] \nonumber\\
	     && +\; \frac{\left[\frac{2 M a}{r} f_{\rm E} +
		\left(1- \frac{2M}{r}\right) f_{\rm L}\right]^2}{
		\Delta \left(1- \frac{2M}{r}\right)} \;.
	\label{kf}
\end{eqnarray}
The derivative of $F$ with respect to $u^r$ is
\begin{eqnarray}
	\frac{\partial F}{\partial u^r} = \frac{2}{u^r}\left[1 + 
		\frac{c_0^2}{r^2 u^r}\left(1-\frac{2M}{r}\right)\right]
		\left[\frac{r^2}{\Delta} \left(u^r\right)^2 -
		\frac{c_0^2}{r^2 u^r}\left(1 - \frac{2M}{r} -
		f_{\rm E}^2\right)\right] \;.
	\label{kdfur}
\end{eqnarray}
The derivative of $F$ with respect to $r$ is too lengthy so we do not 
write it out here. 

The radial velocity at the fast critical point can be solved from $\left.
\partial F/\partial u^r\right|_{r = r_{\rm f}} = 0$
\begin{eqnarray}
	u_{\rm f}^r = - \left[\frac{c_0^2 \Delta_{\rm f}}{r_{\rm f}^4}
		\left( -1 + \frac{2M}{r_{\rm f}} + f_{\rm E}^2\right)
		\right]^{1/3} \;, \hspace{1cm} 
	     \Delta_{\rm f}\equiv \Delta(r = r_{\rm f}) \;.
	\label{kufr}
\end{eqnarray}

From the radial momentum equation $F(r,u^r) = 0$, we can solve for $f_{\rm 
L}$
\begin{eqnarray}
	f_{\rm L} &=& - \frac{1}{1-\frac{2M}{r}} \left\{\frac{2Ma}{r} 
		f_{\rm E} + \left[1 + \frac{c_0^2}{r^2 u^r}\left(1-
		\frac{2M}{r}\right)\right] \right.
		\nonumber\\
		&&\left.\times\left[-\left(1 -\frac{2M}{r}
		\right)\left(r u^r\right)^2 - \Delta \left(1 -
		\frac{2M}{r} - f_{\rm E}^2\right)\right]^{1/2}\right\}
		\;.
	\label{kfl}
\end{eqnarray}
Substituting Eq.~(\ref{kfl}) into the expression for $\partial F/
\partial r$, we can eliminate $f_{\rm L}$ from $\partial F/\partial r$.
Then, similar to the case of a Schwarzschild black hole, we can obtain
$r_{\rm f}$ and $u_{\rm f}^r$ by solving the joined equations (\ref{kufr})
and $\partial F/\partial r = 0$. Then by Eq.~(\ref{kfl}) we can 
determine $f_{\rm L} = f_{\rm L}(c_0, f_{\rm E})$.

For physical solutions $f_{\rm L}$ must be real, so the following condition 
must be satisfied
\begin{eqnarray}
	-\left(1 -\frac{2M}{r}
		\right)\left(r u^r\right)^2 - \Delta \left(1 -
		\frac{2M}{r} - f_{\rm E}^2\right) \ge 0 \;.
	\label{con2a}
\end{eqnarray}
Making use of Eq.~(\ref{kufr}), at $r = r_{\rm f}$ Eq.~(\ref{con2a})
becomes
\begin{eqnarray}
	\left(1- \frac{2M}{r_{\rm f}}\right)^3\, \frac{c_0^4}{r_{\rm f}^2} 
		\le \Delta_{\rm f} \left(-1 + \frac{2M}{r_{\rm f}} + 
		f_{\rm E}^2\right) \;.
	\label{con2}
\end{eqnarray}
As in the case of a Schwarzschild black hole, Eq.~(\ref{con2}) provides 
a constraint on the region in the $(r_{\rm f},c_0)$ space where physical
solutions exist, whence $f_{\rm E}$ is specified. 

In Fig.~\ref{figk1} we show the solutions for $r_{\rm f}$ with $r_0 =
r_{\rm ms}(a)$ and $f_{\rm E} = -E_{\rm Keplerian}(r_0,a)$, where $E_{\rm 
Keplerian}(r_0,a)$ is the specific energy of a test particle moving on a 
Keplerian circular orbit around a Kerr black hole at $r = r_0$ 
\cite{bar72,nov73,pag74}. Each panel corresponds to a different spinning 
state of the black hole. The dashed curve shows the boundary for physical 
solutions: above the dashed curve Eq.~(\ref{con2}) is violated so 
physical solutions do not exist. For the cases of $a/M = 0.95$ and  
$0.998$, we have $r_{\rm ms}< 2M$ and Eq.~(\ref{con2}) is always satisfied
since the right-hand side is always positive [see Eq.~(\ref{kufr}) where
$u_{\rm f}^r < 0$]. So, the 
dashed curve does not appear in the panels corresponding to $a/M = 0.95$ and 
$0.998$. In the case of $a/M = -0.9$, there are two branches of 
solutions for $r_{\rm f}$, labeled by A and B. This is a general feature 
for the case of $a<0$. Like in the case of a Schwarzschild black hole, in the
limit of $c_0^2/r_{\rm g}^2 \rightarrow 0$, $r_{\rm f}$ approaches $r_{\rm ms}$, 
except for the case of $a/M$ very close to $1$, e.g.,  $a/M = 0.998$. In the 
case of $a/M = 0.998$, solutions for $r_{\rm f}$ do not exist if $c_0^2/r_{\rm 
g}^2$ is sufficiently small. This means that for sufficiently small $c_0^2/
r_{\rm g}^2$, the flow is already supersonic as it leaves the marginally
stable circular orbit. This is caused by the fact that we have neglected the 
gas pressure in our treatment. In the Keplerian disk region ($r > r_{\rm ms}$) 
the gas pressure cannot always be neglected---especially near the inner boundary
($r \approx r_{\rm ms}$). When the gas pressure is included, it will make the 
radial motion subsonic in the disk region and give 
rise to a sonic point close to the marginally stable circular orbit 
(\cite{muc82,abr89,afs02}).

In Fig.~\ref{figk2} we show the solutions for $r_{\rm f}$ with $r_0 =
\infty$ and $f_{\rm E} = -1$. Similar to the case of a Schwarzschild black 
hole, when $c_0^2/r_{\rm g}^2\ll 1$ the fast critical point approaches the 
marginally bound circular orbit: $r_{\rm f} \approx r_{\rm mb}$, where
$r_{\rm mb} = 2M \left[1 - a/2M + (1 - a/M)^{1/2}\right]$ \cite{bar72}.

By comparing Figs.~\ref{figk1} and \ref{figk2} with Figs.~\ref{fig1} 
and \ref{fig2}, we can see how the topology of the solutions for $r_{\rm
f}$ evolves with the spin of the black hole. For $a<0$, the solutions have
two branches: A and B. As $a/M$ increases from $-1$ to $0$, the branch
B moves toward left. (One can check this by solving $r_{\rm f}$ for a
negative $a > -0.9$.) As $a$ approaches $0^-$, the branch $B$ of negative
$a$ disappears, the branch $A$ of negative $a$ becomes the branches A and
C of $a = 0$. For $a>0$, the solutions have only one branch. As $a/M$
decreases from $1$ to $0^+$, the branch of positive $a$ becomes the 
branches A and B of $a = 0$. 

As examples, we have solved the radial momentum equation for a Kerr black
hole with $a/M = 0.95$ and $a/M = -0.9$, respectively. We assume that $c_0 = 
0.1\, r_{\rm g}$ and $f_{\rm E} = -1$ (i.e., $r_0 = \infty$). In 
Fig.~\ref{figk3} we show the specific angular momentum of fluid particles 
(upper panel), and the ratio of the electromagnetic energy density to the 
kinetic energy density as measured by a LNRF observer [lower panel, defined 
by Eq.~(\ref{zeta})] for the case of $a/M = 0.95$. In the upper panel, the 
dashed curve shows the corresponding solution for a thin Keplerian disk 
joined to a free-fall flow inside the marginally stable circular orbit. The 
two arrows show the positions of the fast critical point (left) and the 
marginally stable circular orbit (right). In the lower panel, in addition 
to the total $\zeta$ (solid line), we also show the contribution from the 
magnetic field (dashed line) and the electric field (dotted line) separately. 
The dark point on the solid line shows the location of the 
fast critical point. The two circles show the location of the black hole 
horizon. Solutions for the specific angular momentum and the ratio 
of the electromagnetic energy density to the kinetic energy density for the 
case of $a/M = -0.9$ are shown in Fig.~\ref{figk4}. 

From Figs.~\ref{figk3} and \ref{figk4} we see that, similar to the case 
of a Schwarzschild black hole, the specific angular momentum of the fluid 
particles is effectively removed by the magnetic field, though the specific 
energy keeps constant during the motion. However, the specific angular 
momentum of the fluid particles as they reach the horizon of the black hole 
depends on the spin of the black hole, which can be seen by comparing 
Figs.~\ref{figk3} and \ref{figk4} with Fig.~\ref{fig5}.

At large radii the energy of the electromagnetic field is about 
equipartitioned with the kinetic energy of the particles, while near and 
inside the fast critical point the ratio of the electromagnetic energy 
density to the kinetic energy density drops to values $\ll 1$. However, 
unlike the case for a Schwarzschild black hole,
the ratio is not zero on the horizon of the black hole when $a\neq 0$. This
is related to the fact that when $a\neq 0$ the magnetic field is not aligned 
with the velocity field near the horizon (caused by the frame dragging effect,
see Sec.~\ref{sec2}), resulting a nonzero electric field as shown by the 
dotted lines in the two figures (lower panels). On the horizon the ratio of 
the electric energy density to the magnetic energy density approaches $1$, so 
the dashed line and the dotted line end on the same small circle. The 
strength of the electric field drops quickly as the radius increases, and
at large radii (where $E_a^\prime E^{\prime a} \propto r^{-8}$, $B_a^\prime 
B^{\prime a} \propto r^{-1}$) the effect of 
the electric field is negligible. 

The fact that the ratio of the electromagnetic energy density to the kinetic 
energy density is small near and inside the fast critical point, for both the 
Schwarzschild case and the Kerr case, explains why the fast critical point 
approaches the marginally bound circular orbit in the limit of $r_0\gg r_{\rm 
g}$ and $c_0^2/r_{\rm g}^2 \ll 1$, because then $E \approx 1$ and the dynamical 
effects of the electromagnetic field are unimportant near and inside the fast 
critical point. The results support the point that the fast 
critical point behaves like an inner
boundary of the flow, beyond which the dynamical effects of the magnetic field
are important, inside which the dynamical effects of the magnetic field are
negligible and fluid particles free-fall to the black hole.

In Sec.~\ref{sec4.2} we have shown that the specific angular momentum of the 
fluid particles on the horizon of the black hole is given by Eq.~(\ref{lh}). 
For solutions that smoothly pass the fast critical point, the parameter $f_{\rm 
L}$ is determined as a function of $f_{\rm E}$ and $c_0$ when $\Psi = 0$. Hence, 
if $f_{\rm E}$ is specified, for a given spin of the black hole $L_{\rm H}$ 
depends only on $c_0$. In Fig.~\ref{figk5}, we show $L_{\rm H}$ as a function 
of $c_0$, where different curves correspond to different spinning states of 
the black hole. The outer boundary of the flow is assumed to be at
infinity so that $f_{\rm E} = -1$. Figure~\ref{figk5} confirms the results in 
Figs.~\ref{fig5}, \ref{figk3}, and \ref{figk4}. Furthermore, we see that 
$L_{\rm H}$ depends on $a$ in opposite ways for small and large $c_0$: for 
small $c_0$, $L_{\rm H}$ decreases with increasing $a$; for large $c_0$, 
$L_{\rm H}$ increases with increasing $a$. And, for sufficiently large $c_0$, 
$L_{\rm H}$ becomes zero when $a = 0$ (see Sec.~\ref{sec5.1}). 
For large $c_0$ and negative $a$, $L_{\rm H}$ is negative. The $L_{\rm H}$ 
corresponding to positive $a$ is always positive.

\section{Discussion: The Effect of Finite Resistivity and Different Evolution
of Large and Small-Scale Magnetic Fields}
\label{sec6}

The solutions that we have found have the following unique feature: the 
magnetic field is aligned with the velocity field [Eq.~(\ref{parel})]; and, 
as a result, 
at large radii the energy of the magnetic field is equipartitioned with
the kinetic energy of the flow [Eq.~(\ref{equi})]. The latter is obvious from 
the following consideration: At large radii, where the gravity is approximately 
Newtonian and the radial flux of angular momentum is not important, we have 
$4\pi \rho_{\rm m} u^r u^\phi \approx B^r B^\phi$. Then, $B^r /B^\phi = u^r 
/u^\phi$ implies that $B_r B^r \approx 4\pi \rho_{\rm m} u_r u^r$ 
and $B_\phi B^\phi \approx 4 \pi \rho_{\rm m} u_\phi u^\phi$, i.e., $B^2 /8\pi 
\approx \rho_{\rm m} v^2 /2$, where $B^2 = B_r B^r + B_\phi B^\phi$ and $v^2 = 
u_r u^r + u_\phi u^\phi$. An almost inverse statement is also true: If at large 
radii $B_\phi B^\phi \approx 4 \pi \rho_{\rm m} u_\phi u^\phi$, then we must 
have $B^r/B^\phi\approx u^r / u^\phi$ at large radii. (See App.~\ref{appb} for
the corresponding solutions for a Newtonian accretion disk.)

The above results rely on the fact that we have neglected all energy dissipation 
processes by assuming a perfectly conducting fluid with zero resistivity and 
viscosity. In the Newtonian limit (which is valid at large distances from the 
central black hole), the induction equation including a finite electric 
resistivity $\eta$ is given by (see, e.g., \cite{bal98})
\begin{eqnarray}
     \frac{\partial{\bf B}}{\partial t} = \nabla\times\left({\bf v}\times
	     {\bf B} - \frac{\eta}{4\pi} \nabla\times{\bf B}\right) \;.
	\label{ind}
\end{eqnarray} 
We assume that $\eta = \mbox{constant}$. Then, substitute Eq.~(\ref{velo}) into 
Eq.~(\ref{ind}), we obtain the stationary and axisymmetric solution
\begin{eqnarray}
     r\left(v_r B_\phi - v_\phi B_r\right) = \frac{\eta}{4\pi}\frac{d}{dr}
		\left(r B_\phi\right) + C \;, \label{rsol1}
\end{eqnarray}
where $C$ is an integral constant. 

To estimate the effect of finite resistivity, let us consider the case of $C = 
0$. Then from Eq.~(\ref{rsol1}) we have
\begin{eqnarray}
     \frac{B_r}{B_\phi} = \zeta_{\rm M}\,\frac{v_r}{v_\phi} \;,
	     \hspace{1cm} \zeta_{\rm M}\equiv 1 -\frac{\eta}
		{4\pi r v_r} \frac{d\ln\left(r B_\phi\right)}
		{d\ln r} \;.
	\label{brfv2}
\end{eqnarray}
The parameter $\zeta_{\rm M}$ determines the orientation of the magnetic
field relative to the velocity field in the $r-\phi$ plane. Define
\begin{eqnarray}
     R_{\rm M} \equiv \frac{8\pi^2 r \left|v_\phi\right|}{\eta} \;,
	     \hspace{1cm} q \equiv \frac{d\ln\left(r B_\phi\right)}
		{d\ln r} \;,
	\label{rbp}
\end{eqnarray}
where $R_{\rm M}$ is the magnetic Reynolds number associated with the azimuthal 
motion, $q$ measures the radial gradient of the magnetic field. Then, 
$\zeta_{\rm M}$ can then be rewritten as
\begin{eqnarray}
     \zeta_{\rm M} = 1 + \frac{1}{\lambda} \;, \hspace{1cm}
	     \lambda \equiv \frac{R_{\rm M}}{2\pi q} \left|\frac{v_r}{v_\phi}
	     \right| \;.
	\label{lambda}
\end{eqnarray}

Therefore, the effect of finite resistivity is determined by the parameter 
$\lambda$. When $|\lambda| \gg 1$, we have $\zeta_{\rm M} \approx 1$, the 
effect of resistivity is unimportant and the solutions correspond to a perfectly 
conducting fluid as in the case of our analytical model. Then, the magnetic 
field is aligned with the velocity field and the magnetic energy density is 
equipartitioned with the rotational energy density of the disk at large radii. 
When $|\lambda| \ll 1$, we have $|\zeta_{\rm M}| \approx 1/|\lambda| \gg 1$,
the effect of resistivity becomes important and the solutions correspond to a 
nonperfectly conducting fluid. The magnetic field is not aligned with the 
velocity field, resulting significant energy transportation and dissipation. 
Then, in the stationary state the magnetic energy density should be much smaller 
than the rotational energy density of the disk (as shown bellow). The effect of 
finite resistivity is particular important for small-scale and chaotic magnetic 
fields that reverse directions frequently, since at places where $B_\phi =0$ but 
$dB_\phi/dr \neq 0$ we have $|q| = \infty$ [Eq.~(\ref{rbp})] and $\lambda =0$.

It is the assumption of a perfectly conducting fluid with zero resistivity and
viscosity that makes our analytical results different from the MHD numerical 
simulation results, where the magnetic field is not always aligned with the 
velocity field and the energy density of the magnetic field is usually much 
smaller than the rotational energy density of disk particles (see, e.g., 
\cite{sto96,haw96,haw00,haw01,haw02,ste02}). In numerical simulations, finite 
resistivity and energy dissipation arising from finite spatial gridding in 
numerical codes always exist and are often large, even though the physical 
resistivity 
and viscosity can be assumed to be zero \cite{haw95,ryu95}. In fact, in current 
three-dimensional MHD simulations, the Reynolds number arising from finite 
numerical gridding is often smaller than the magnetic Reynolds number associated 
with most astrophysical plasmas \cite{haw95,haw01a}.

How about the effect of finite resistivity in real astronomy? For astrophysical 
plasmas, the magnetic Reynolds number $R_{\rm M}$ is usually extremely large
($\agt 10^6$) due to small microscopic Ohmic resistivity and large spatial
scales and velocities (see, e.g., \cite{cho98}). However, dissipation processes 
like convection, turbulence, etc, can induce a large effective resistivity
\cite{kra80,pud81a,pud81b,mes99}, thus produce a smaller magnetic Reynolds 
number. The parameter $q$ can also be large if the magnetic field has small scales 
and is chaotic. 

For an ordered large-scale magnetic field we usually have $q \sim 1$ in regions 
where $B_\phi\neq 0$. While for small-scale and chaotic magnetic fields, the 
parameter $q$ can be very large, especially at places where $B_\phi= 0$. In
this latter case, the accretion disk cannot always be described with a simple 
stationary and axisymmetric model without using a statistical approach. We can
define a coherent length scale $l_{\rm c}$ so that $|q| \sim r / l_{\rm c}$.
Then, $\lambda$ in Eq.~(\ref{lambda}) can be recast as
\begin{eqnarray}
     \lambda \sim \frac{R_{\rm c}}{2\pi} \left|\frac{v_r}{v_{\rm c}}\right| 
	     \;, \hspace{1cm} 
		R_{\rm c} \equiv \frac{8\pi^2 l_{\rm c} v_{\rm c}}{\eta} \;,
	\label{lam2}
\end{eqnarray}
where $v_{\rm c}$ is the velocity associated with the local coherent structure,
$R_{\rm c}$ is the corresponding magnetic Reynolds number, and $\eta$ is the 
resistivity in the disk.

Thus, accretion disks driven by large-scale magnetic fields behave very 
differently from those driven by small-scale magnetic fields. For an accretion
disk driven by a large-scale magnetic field, like the model studied in this 
paper, $\lambda$ is usually $\gg 1$ (since $q\sim 1$) and determined by the 
global motion of the disk through $R_{\rm M}$ and $v_r/v_\phi$ 
[Eq.~(\ref{lambda})]. For an accretion disk driven by small-scale magnetic 
fields, $\lambda$ is usually $\ll 1$ (since $|q|\gg 1$) and determined by 
$R_{\rm 
c}$ and $v_r/v_c$ [Eq.~(\ref{lam2})]. In this latter case, the global motion 
affects $\lambda$ (so affects the equilibrium state) only through the radial 
velocity $v_r$---the global rotation of the disk has no effect on $\lambda$
(so does not affect the equilibrium state of the magnetic field).

Making use of Eq.~(\ref{brfv2}) and $4\pi\rho_{\rm m} v_r v_\phi \approx B_r 
B_\phi$, we can estimate the energy density of the magnetic field by
\begin{eqnarray}
     \frac{{\cal E}_{\rm M}}{{\cal E}_{\rm K}} \approx
	   \frac{\zeta_{\rm M}^{-1} + \zeta_{\rm M} \left(v_r/v_\phi\right)^2}
	   {1 + \left(v_r/v_\phi\right)^2} \;,
\end{eqnarray}
where ${\cal E}_{\rm M} = (1/8\pi) \left(B_r^2 + B_\phi^2\right)$ is the magnetic 
energy density, ${\cal E}_{\rm K} = \left(\rho_{\rm m}/2\right) \left(v_r^2 + 
v_\phi^2\right)$ is the kinetic energy density of the disk. When $\lambda \gg 1$
(corresponding to large-scale magnetic fields), we have ${\cal E}_{\rm M} 
\approx {\cal E}_{\rm K}$, the magnetic energy is approximately equipartitioned 
with the rotational energy of the disk (recall that $\left|v_r/v_\phi\right| \ll 
1$). When $\lambda \ll 1$ (corresponding to small-scale magnetic fields), we 
have ${\cal E}_{\rm M} / {\cal E}_{\rm K} \sim \left(R_{\rm c}/2\pi\right)\left|
v_r/v_\phi\right| \sim \lambda \left|v_{\rm c}/v_\phi\right|\ll 1$, the magnetic 
energy density is much smaller than the rotational energy density of the disk. 
As an example for an accretion disk with
small-scale magnetic fields, consider a thin Keplerian disk with $\left|v_r
\right|\sim v_{\rm c}^2/\left|v_\phi\right|$ where $v_{\rm c}$ corresponds to
the fluctuation in the fluid velocity arising from turbulence \cite{bal98}, 
and $R_{\rm c} \sim 2\pi$ (see, e.g., \cite{mes99}), then we have ${\cal 
E}_{\rm M} \sim \rho_{\rm m} v_{\rm c}^2/2$, i.e., the magnetic energy density 
is approximately equipartitioned with the kinetic energy density in turbulence. 
Thus, for a disk with small-scale magnetic fields the equilibrium state is 
determined by the local turbulent motion rather than the global rotation
\cite{pud81a,pud81b}. 

Based on the above arguments, we believe that the results of Gammie \cite{gam99}
are more relevant to a large-scale magnetic field connecting the disk to the 
black hole, rather than small-scale and chaotic magnetic fields in the plunging
region.

\section{Summary and Conclusions}
\label{sec7}

With an analytical model we have investigated the dynamics of a cold accretion 
disk containing a large-scale magnetic field around a Kerr black hole, trying to 
understand the process of angular momentum and energy transportation. A 
one-dimensional radial momentum equation is derived near the equatorial plane 
of the black hole, which has one intrinsic singularity at the fast critical 
point. Any physical solution corresponding to an accretion flow starting from 
infinity subsonically must 
smoothly pass the fast critical point in order to reach the horizon of the black 
hole. Because of this requirement, among the four integral constants ($c_0$, 
$\Omega_\Psi$, $f_{\rm L}$, and $f_{\rm E}$) only three are independent: one of 
them has to be determined as a function of the other three.

In order for $r^2 B^r$ and $B^2/4\pi\rho_{\rm m}$ to be finite at infinity, the 
constant $\Omega_\Psi$ has to be zero (Secs.~\ref{sec2} and \ref{sec3.1}). 
This indicates that in the stationary state the magnetic field and the velocity 
field of the fluid must be parallel to each other. When this 
condition is satisfied ($\Omega_\Psi= 0$), the specific energy of the fluid 
particles remains constant during the motion, but the specific angular 
momentum changes. 

\begin{figure}
\vspace{3cm}
\includegraphics[width=15.5cm]{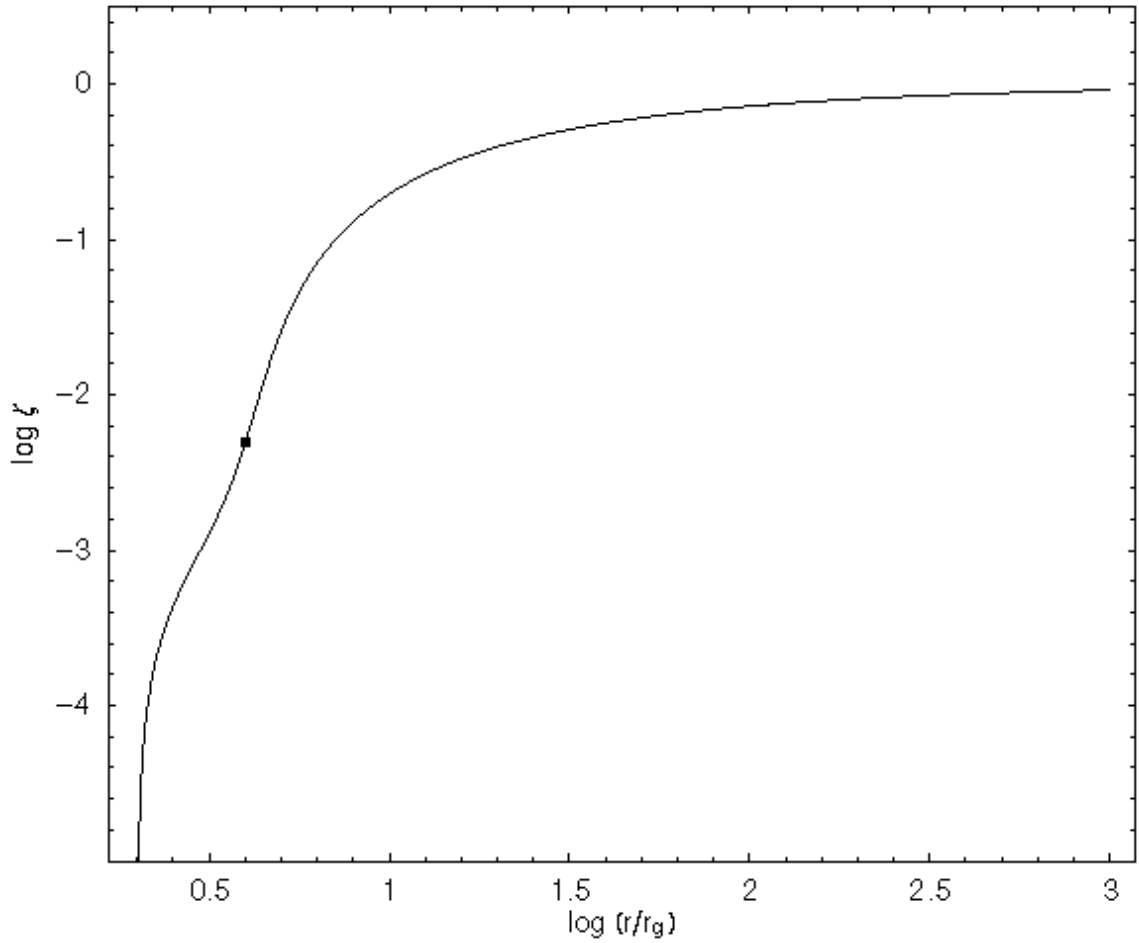}
\caption{The ratio of the magnetic energy density to the kinetic energy density
[defined by Eq.~(\ref{zeta})] corresponding to the solution in Fig.~\ref{fig4}. 
The dark point shows the location of the fast critical point.
\label{fig7}}
\end{figure}

\begin{figure}
\vspace{3cm}
\includegraphics[width=15.5cm]{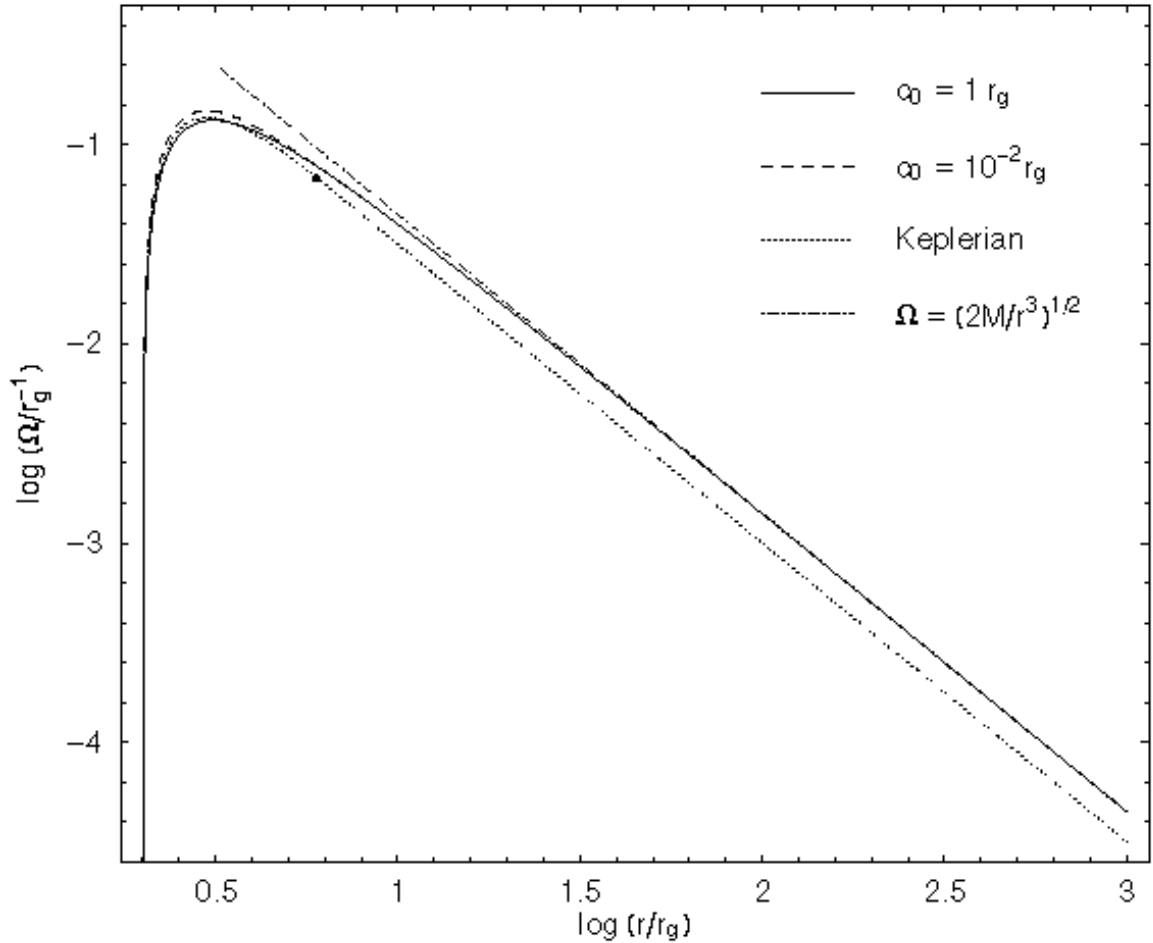}
\caption{The angular velocity of a flow with $f_{\rm E} = -1$. The parameter 
$c_0$ is alternatively assumed to be $1\, r_{\rm g}$ (solid line) and $10^{-2}
\, r_{\rm g}$ (dashed line). The dotted line shows the angular velocity of a 
thin Keplerian disk joined to a free-fall flow inside the marginally stable 
circular orbit (the dark point). The dashed-dotted line 
represents the asymptotic solution given by Eq.~(\ref{uphiasp}).
\label{fig8}}
\end{figure}

\begin{figure}
\vspace{1cm}
\includegraphics[width=15.5cm]{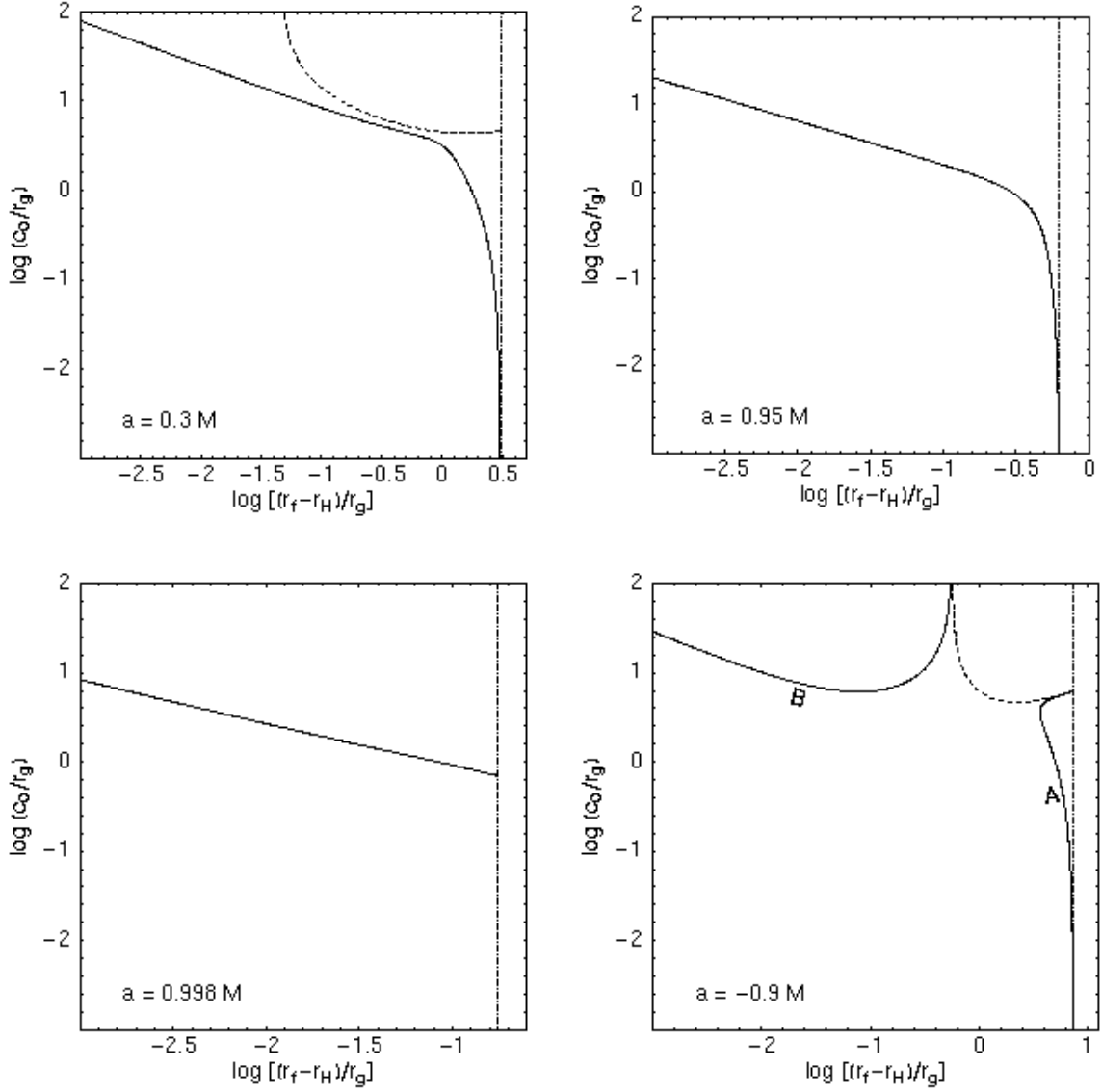}
\caption{Solutions for $r_{\rm f}$---the radius at the fast critical 
point corresponding to a flow with constant specific energy accreting onto 
a Kerr black hole (solid curves). Each panel corresponds to a different 
spinning state of the black hole, as indicated by the value of $a$. The 
outer boundary of the flow is at the marginally stable circular orbit: $r_0 
= r_{\rm ms}(a)$, as indicated by the vertical dotted line. The specific 
energy of fluid particles is assumed to be equal to that of a test particle 
moving on a Keplerian circular orbit at $r_0 = r_{\rm ms}(a)$. The dashed 
curve represents the boundary for physical 
solutions: beyond the dashed curve physical solutions do not exist (see 
the text). 
\label{figk1}}
\end{figure}

\begin{figure}
\vspace{2cm}
\includegraphics[width=15.5cm]{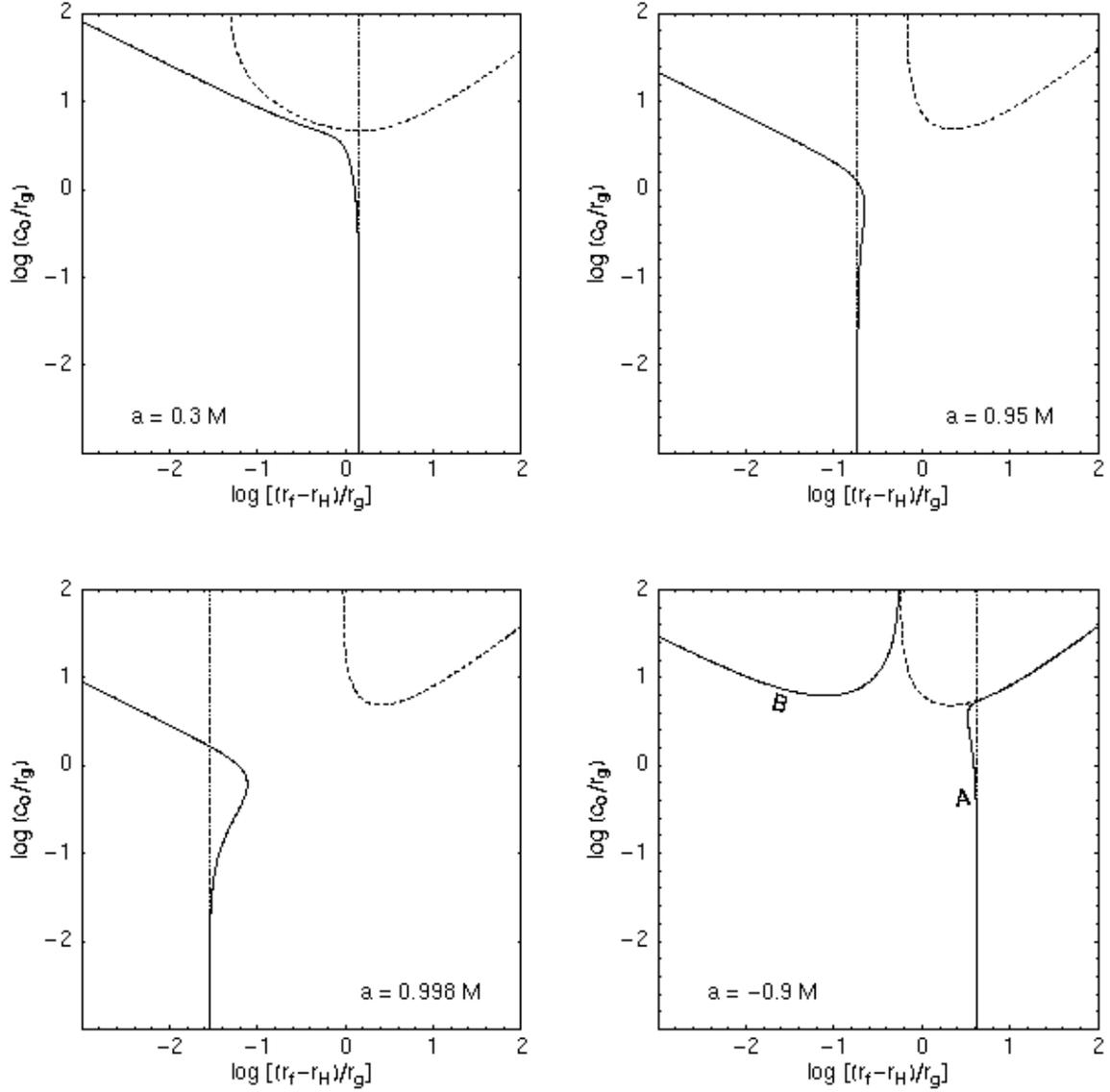}
\caption{Same as Fig.~\ref{figk1} but $r_0 = \infty$. The specific energy
of fluid particles is assumed to be equal to $1$. The vertical dotted line 
marks the location of the marginally bound circular orbit, $r_{\rm mb}(a)$.
\label{figk2}}
\end{figure}

\begin{figure}
\includegraphics[width=12.8cm]{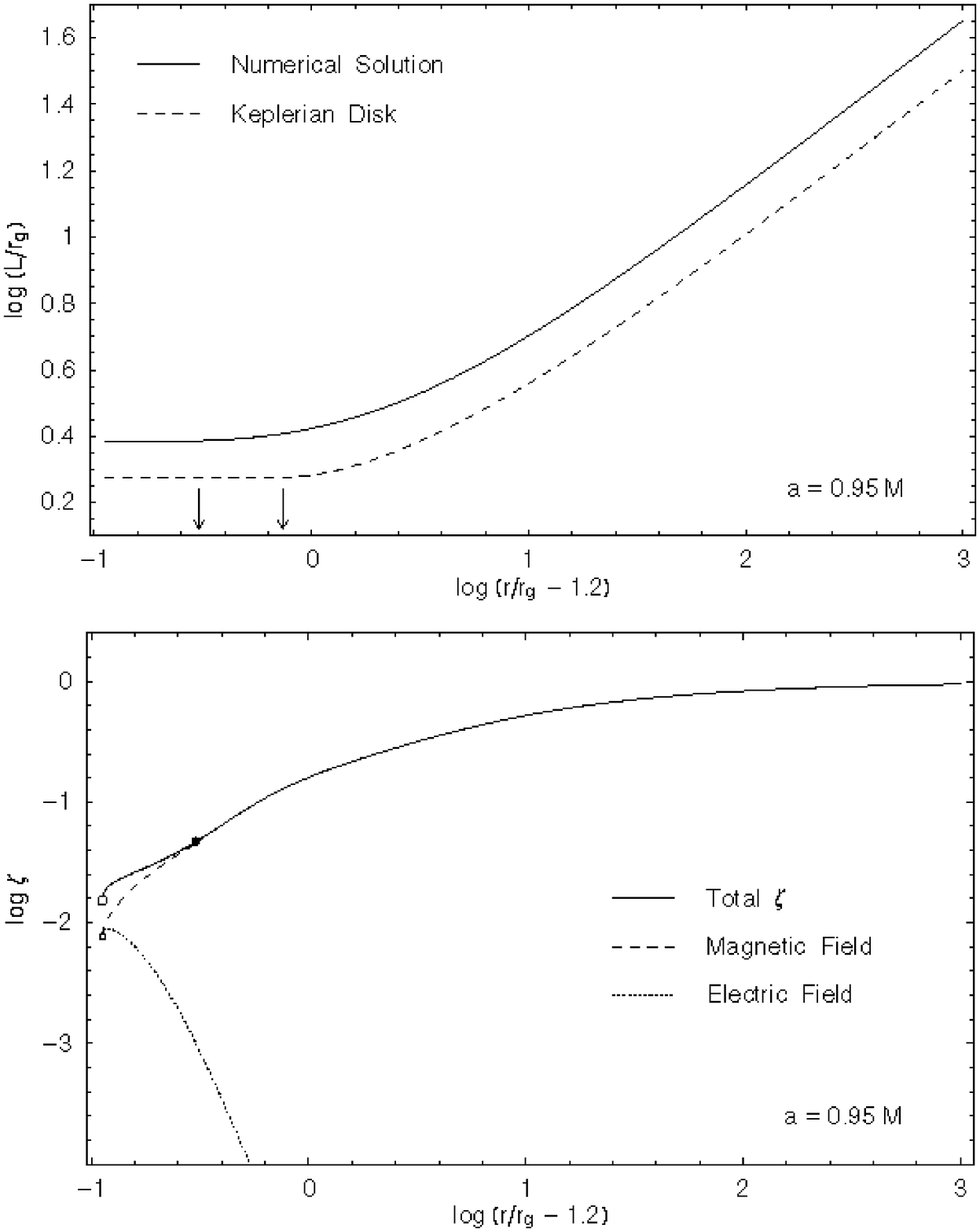}
\caption{Upper panel: Evolution of the specific angular momentum of a flow 
with constant specific energy accreting onto a Kerr black hole with $a/M = 
0.95$. The dashed curve shows the solution corresponding to a thin Keplerian 
disk joined to a free-fall flow inside the marginally stable circular orbit. 
The two arrows show the locations of the fast critical point (left) and the 
marginally stable circular orbit (right). Lower panel:  
Ratio of the electromagnetic energy density to the kinetic energy density 
(solid line) corresponding to the solution in 
the upper panel, which consists of the contributions of magnetic field
(dashed line) and electric field (dotted line). The dark point shows the 
location of the fast critical point. The circles show location of the horizon. 
(The unit of the horizontal axis is chosen to make the structure near the 
horizon clear.)
\label{figk3}}
\end{figure}

\begin{figure}
\vspace{2cm}
\includegraphics[width=14.4cm]{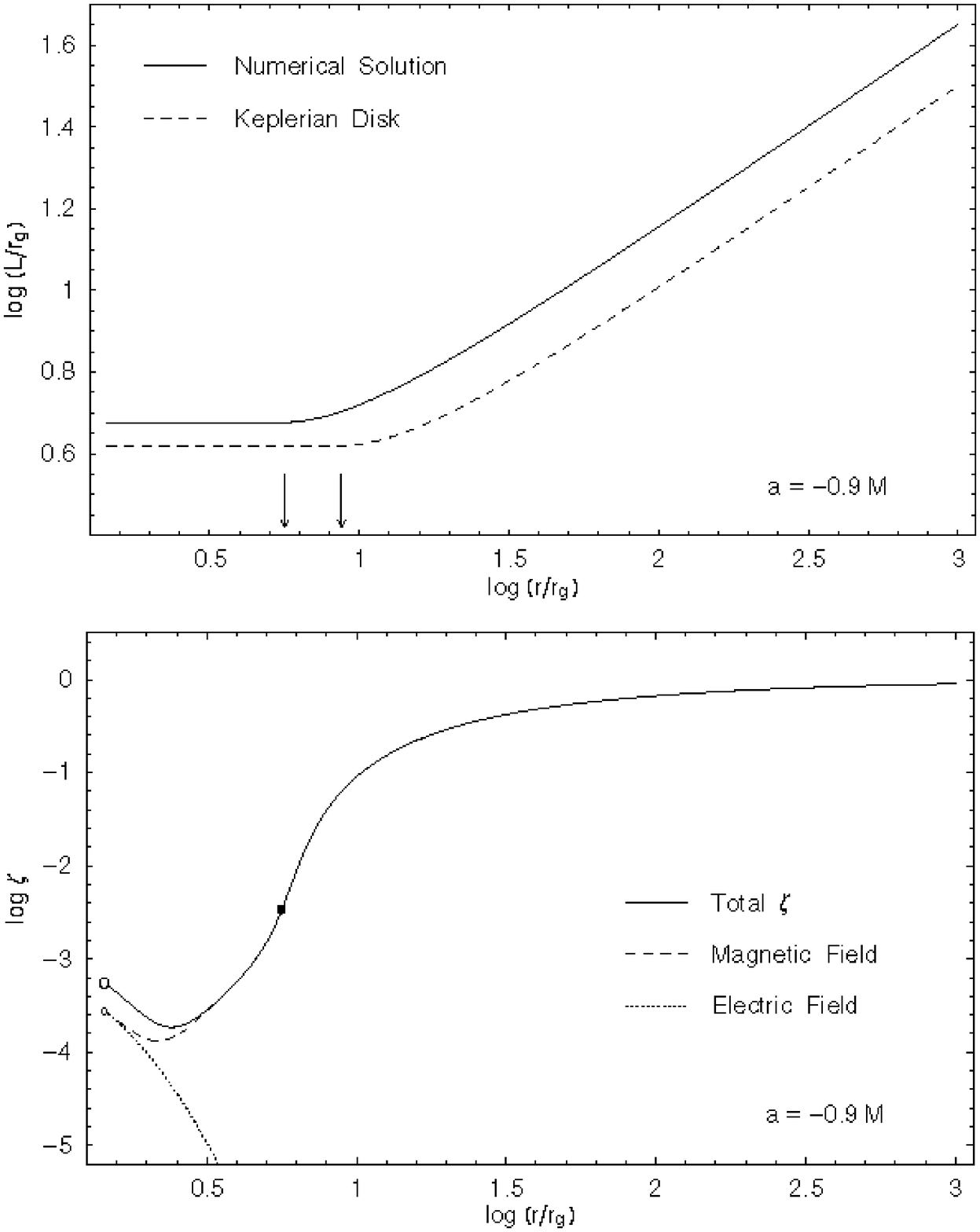}
\caption{Same as Fig.~\ref{figk3} but for $a/M = -0.9$.
\label{figk4}}
\end{figure}

\begin{figure}
\vspace{2cm}
\includegraphics[width=15.5cm]{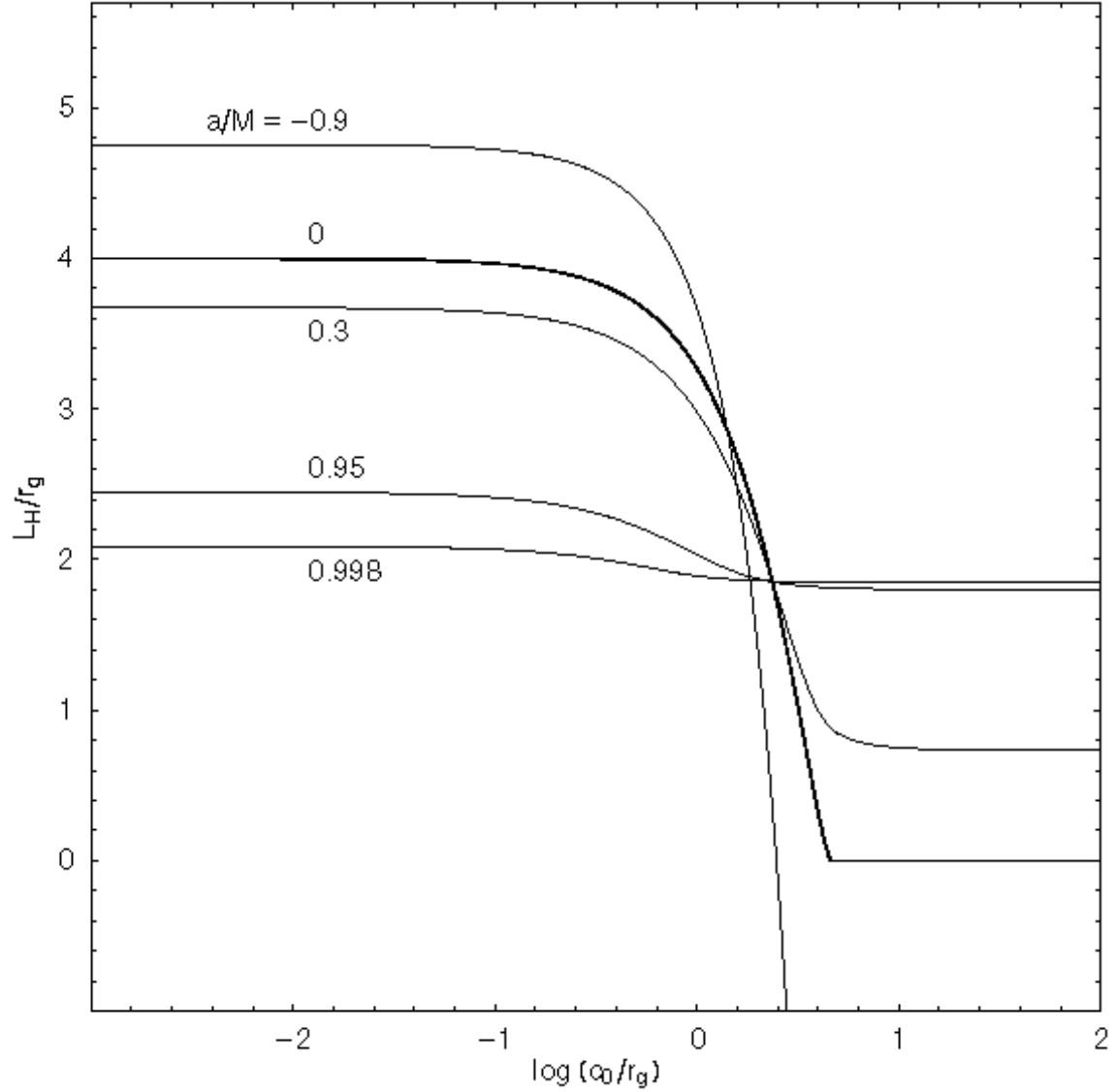}
\caption{The specific angular momentum of fluid particles as they reach 
the horizon of the black hole, as a function of $c_0$. Different curves 
correspond to different spinning states of the black hole as labeled. The 
outer boundary of the flow is at $r_0 = \infty$ so $f_{\rm E} = -1$. (At 
large radii the specific angular momentum is $\propto r^{1/2}$, so $L = 
\infty$ at $r_0 = \infty$.) 
\label{figk5}}
\end{figure}

The radial momentum equation has been solved with the assumptions that 
$\Omega_\Psi = 0$, $c_0^2/r_{\rm g}^2 \ll 1$, and $f_{\rm E} = -1$ (corresponding 
to an accretion flow with the outer boundary at infinity). The results can be 
summarized as follows:
\begin{itemize}
\item[1.]{The accretion flow starts from a subsonic state at large
radii, passes the fast critical point near the marginally bound circular
orbit, then enters the black hole supersonically (see, e.g., Figs.~\ref{fig4}).}
The solutions are super-Keplerian (Figs.~\ref{fig5}, \ref{fig8}, \ref{figk3} and 
\ref{figk4} upper panels), whose asymptotic behaviors on the black 
hole horizon and at infinity are summarized in Table~\ref{tab1}. 
\item[2.]{The specific energy of the fluid particles remains constant, but
the specific angular momentum is effectively removed by the magnetic field and 
transported outward (Figs.~\ref{fig5}, \ref{figk3} and \ref{figk4} upper panels). 
The rate of transporting angular momentum is similar to that for a thin Keplerian 
disk, though the specific angular momentum that the flow finally carries into the 
black hole is somewhat larger.}
\item[3.]{At large radii the electromagnetic energy density is about 
equipartitioned with the kinetic energy density of the fluid particles, but
near and inside the fast critical point the ratio of the electromagnetic energy 
density to the kinetic energy density is $\ll 1$ (Figs.~\ref{fig7}, \ref{figk3} 
and \ref{figk4} lower panels). This is due to the fact that at large radii the 
velocity of the fluid is predominantly toroidal so the magnetic field is 
sufficiently amplified by the shear rotation, while near and inside the fast 
critical point the radial velocity of the fluid becomes important then the 
magnetic field has no time to get sufficient amplification 
(see Fig.~\ref{fig6}).}
\item[4.]{When the black hole is Schwarzschild, the electric field measured
by a local static observer is zero and the ratio of the magnetic energy density
to the kinetic energy density approaches zero on the horizon of the black
hole (Fig.~\ref{fig7}).}
\item[5.]{When the black hole is Kerr, due to the frame dragging effect the 
electric field measured by a LNRF observer is nonzero. The electric field decays
rapidly with increasing radius, but near the horizon the electric field makes a 
significant contribution to the total electromagnetic energy density. On the 
horizon, the ratio of the total electromagnetic energy density to the kinetic 
energy density, which is contributed equally by electric field and magnetic field, 
is tiny but nonzero (see Figs.~\ref{figk3} and \ref{figk4}}, lower panels).
\item[6.]{The specific angular momentum of the fluid particles as they reach 
the horizon, $L_{\rm H}$, is a strong function of $a/M$ and $c_0/r_{\rm g}$, 
though in the limit $c_0^2/r_{\rm g}^2 \ll 1$ $L_{\rm H}$ becomes insensitive 
to the change in $c_0/r_{\rm g}$ (Fig.~\ref{figk5}). For a positive $a$, 
$L_{\rm H}$ is always positive. For a zero or negative $a$, $L_{\rm H}$ becomes 
zero or negative for sufficiently large $c_0/r_{\rm g}$.}
\end{itemize}

With the existence of solutions with constant specific energy for a magnetized
accretion flow, the question posed in the Introduction ({\it Do magnetic fields 
transport or dissipate energy as efficiently as they transport angular 
momentum?}) is at least partly answered: {\it Magnetic 
fields do not have to transport or dissipate a lot of energy as they 
efficiently transport angular momentum.} This has important implications to 
the theory of accretion disk: an accretion disk driven by magnetic fields 
may have a very low radiation efficiency. People generally believe that an 
accretion disk usually has a radiation efficiency much higher than a 
spherical accretion flow. The belief comes from the following argument: a 
disk has a lot of angular momentum which must be removed in order for the 
disk material to be able to accrete onto the central compact object. In
the meantime, a lot of energy is expected to be removed also. The results in 
this paper show that the specific energy of fluid particles does not have to 
change when its angular momentum is being removed. In fact the solutions with 
constant specific energy are the unique solutions for the model studied in 
this paper, all other solutions have bad 
asymptotic behaviors at infinity (namely, $r^2 B^r$ and $B^2/4\pi\rho_{\rm 
m}$ diverge), if the specific angular momentum of the fluid is unbounded 
at infinity. 

We have ignored energy dissipation by assuming that the disk 
plasma is perfectly conducting. Then, inevitably, at large radii the magnetic 
energy density is equipartitioned with the kinetic energy density, indicating 
that the accretion flow must be geometrically thick. The solutions presented in 
the text apply only to the region near the central plane of the disk. However, 
in App.~\ref{appa} we show that the solutions can be easily generalized to more 
general and more complicated models, including the disk region well above or 
well below the equatorial plane. So, solutions with constant specific energy 
exist for more general and more complicated models.

The effects of finite resistivity are discussed in Sec.~\ref{sec6}. When the 
resistivity is 
nonzero, an accretion disk driven by a large-scale and ordered magnetic field 
behaves very differently from that driven by small-scale and chaotic magnetic 
fields. For a large-scale and ordered magnetic field, in a stationary and 
axisymmetric state the magnetic energy density is approximately equipartitioned 
with the rotational energy density at large radii if the resistivity is small. 
For small-scale and chaotic magnetic fields that reverse directions frequently, 
even if the resistivity is small, in the equilibrium state the magnetic energy 
density is likely to be equipartitioned with the kinetic energy density 
associated with local random motions (e.g., turbulence). 

A finite resistivity in the disk will give rise to a nonzero radiation 
efficiency. However, a nonzero radiation efficiency may also arise in the 
following way: a disk corresponding to our solutions has a finite outer radius 
$r_0$ and beyond which it is joined to a thin Keplerian disk. Then, the 
arguments in Sec.~\ref{sec3.1} indicate that the region 
inside $r_0$ might have a radiation efficiency $\varepsilon_{\rm in}
\alt \Omega_0 L_0 \sim r_{\rm g}/r_0\,$, assuming that $r_0\gg r_{\rm g}\,$. 
The thin Keplerian disk outside $r_0$ would give rise to a radiation efficiency 
$\varepsilon_{\rm out} \sim r_{\rm g}/r_0\,$. Then, the total radiation
efficiency of the disk region from $r = r_{\rm H}$ to $r = r_\infty$ would
be $\varepsilon_{\rm total} = \varepsilon_{\rm in} + \varepsilon_{\rm out}
\sim r_{\rm g}/r_0\,$.

Finally, we remark that the issue of stability/instability has been ignored in
our treatment. MHD instabilities are often important in the study of magnetized 
accretion disks (for reviews see \cite{bal98,mes99,bla02}), they may also be 
important for our solutions. However, a detail treatment of the 
stability/instability of our solutions is beyond the scope of the current paper, 
so we would leave it for future work.

\begin{acknowledgments}
The author thanks Steven Balbus, Ramesh Narayan, and Bohdan Paczy\'{n}ski for 
advices and discussions. This research was supported by NASA through
Chandra Postdoctoral Fellowship grant number PF1-20018 awarded by the
Chandra X-ray Center, which is operated by the Smithsonian
Astrophysical Observatory for NASA under contract NAS8-39073.
\end{acknowledgments}

\appendix
\section{Generalization of the Solutions to More General Models}
\label{appa}

The results presented in the text can be generalized to more general
models where the stream lines of the fluid do not lie on surfaces of
constant $\theta$. To do so, let us assume that the disk can be foliated 
with a set of surfaces so that each stream line of the flow lies on one 
of such surfaces. We call the surfaces so defined the {\it stream 
foliation surfaces}. By symmetry, the equatorial plane is one of the
stream foliation surfaces. The upper and lower faces of the disk are two 
stream foliation surfaces.

For a stationary and axisymmetric disk, the stream foliation surfaces
are specified by some functions of $r$ and $\theta$. We can define a
congruence of lines that perpendicularly intersect each stream
foliation surface. Each point on the equatorial plane has a radial
coordinate $r$, the congruence of lines carry this radial coordinate
to each of the other stream foliation surfaces.  We label each stream 
foliation surface with a variable $\Theta$, and choose $\Theta = 0$ on 
the equatorial plane, $\Theta = \pm 1$ on the upper and lower faces
of the disk. Then, we are able to define two new coordinates $R$ and 
$\Theta$, which are related to $r$ and $\theta$ by
\begin{eqnarray}
	R = R(r,\theta) \;, \hspace{1cm}
	\Theta = \Theta(r, \theta) \;,
	\label{rth1}
\end{eqnarray}
so that each stream foliation surface corresponds to $\Theta =
\mbox{constant}$ and the coordinate lines of $R$ and $\Theta$ are
orthogonal to each other. On the equatorial plane $\Theta = 0$ we 
have $R = r$. The relations in Eq.~(\ref{rth1}) should give rise to a 
one-to-one correspondence between $(r,\theta)$ and $(R,\Theta)$ at least 
locally. 

A general stationary and axisymmetric metric can be written as
\begin{eqnarray}
	ds^2 = g_{tt}\, dt^2 + 2 g_{t\phi}\, dt d\phi + g_{\phi\phi}\, 
		d\phi^2+ g_{rr}\, dr^2 + g_{\theta\theta}\, d\theta^2 \;,
	\label{met1}
\end{eqnarray}
where $g_{\mu\nu} = g_{\nu\mu}$ are functions of $r$ and $\theta$. Now let
us make coordinate transformations from $(t,r,\theta,\phi)$ to $(t,R,\Theta,
\phi)$. Then the metric becomes
\begin{eqnarray}
	ds^2 = g_{tt}\, dt^2 + 2 g_{t\phi}\, dt d\phi + g_{\phi\phi}\, 
		d\phi^2+ g_{RR}\, dR^2 + g_{\Theta\Theta}\, d\Theta^2 \;,
	\label{met2}
\end{eqnarray}
where 
\begin{eqnarray}
	g_{RR} = g_{rr} \left(\frac{\partial r}{\partial R}\right)^2 
		+ g_{\theta\theta} \left(\frac{\partial \theta}
		{\partial R}\right)^2 \;, \hspace{1cm}
	g_{\Theta\Theta} = g_{rr} \left(\frac{\partial r}{\partial 
		\Theta}\right)^2 + g_{\theta\theta} \left(\frac{\partial 
		\theta}{\partial \Theta}\right)^2 \;.
\end{eqnarray} 
The metric components $g_{\mu\nu}$ in Eq.~(\ref{met2}) are functions
of $R$ and $\Theta$. 

In deriving Eq.~(\ref{met2}) we have used
\begin{eqnarray}
	g_{rr}\frac{\partial r}{\partial R} \frac{\partial r}{\partial 
		\Theta} + g_{\theta\theta} \frac{\partial \theta}
		{\partial R}\frac{\partial \theta}{\partial \Theta}
		= 0 \;,
\end{eqnarray}
since the coordinate lines of $\Theta$ are orthogonal to the stream foliation 
surfaces. 

In coordinates $(t,R,\Theta,\phi)$, the Maxwell equation to be solved is 
again given by Eq.~(\ref{maxeq}), but now we have $x^{\alpha} = (t,R,
\Theta,\phi)$ and 
\begin{eqnarray}
	g = g_{RR} g_{\Theta\Theta} \left(g_{tt} g_{\phi\phi}
		- g_{t\phi}^2\right) \;.
\end{eqnarray}
Assuming that each magnetic field line lies in one of the stream foliation
surfaces, then
\begin{eqnarray}
	u^\Theta = B^\Theta = 0 \;, \hspace{1cm}
	\frac{\partial u^\Theta}{\partial \Theta} =
		\frac{\partial B^\Theta}{\partial \Theta}
		= 0 \;.
	\label{pur}
\end{eqnarray}
Equation~(\ref{pur}) is the generalization of Eqs.~(\ref{bth}) and 
(\ref{dbth}). This assumption should not limit our results too much, since 
the $\Theta$-component of the magnetic field is not relevant to the 
transportation of angular momentum in the $R$-direction.

Then, for stationary and axisymmetric solutions the Maxwell equation
becomes
\begin{eqnarray}
	\frac{\partial}{\partial R} \left[\sqrt{-g}\left(u^R 
		B^\beta - u^\beta B^R \right)\right] = 0 \;.
	\label{maxa1}
\end{eqnarray}
Making use of $u_a B^a = 0$ and $u_a u^a = -1$, we obtain the solutions to 
Eq.~(\ref{maxa1})
\begin{eqnarray}
	B^R &=& \frac{1}{\sqrt{-g}} \left(-C_0 u_t + \Psi 
		u_\phi\right) \;, 
		\label{bra} \\
 	B^\phi &=& \frac{1}{\sqrt{-g}\, u^R} \left[-C_0 u_t u^\phi
		+ \left(1+ u_\phi u^\phi\right)\Psi\right] \;, 
		\label{bfa} \\
	B^t &=& \frac{1}{\sqrt{-g}\, u^R} \left[- \left(1+ u_t 
		u^t\right)C_0+ u_\phi u^t\Psi\right] \;,
		\label{bta}
\end{eqnarray}
where $C_0 = C_0 (\Theta)$, $\Psi = \Psi(\Theta)$. From Eqs.~(\ref{bra}) 
and (\ref{bfa}) we have 
\begin{eqnarray}
	\frac{B^R}{B^\phi} = \frac{u^R}{u^\phi} 
\end{eqnarray}
when $\Psi = 0$.

Substituting the solutions~(\ref{bra})--(\ref{bta}) into Eq.~(\ref{tab}), we 
obtain the $\phi-R$ and $t-R$ components of the electromagnetic 
stress-energy tensor
\begin{eqnarray}
        T_{{\rm EM},\phi}^{~~~~~R} = - \frac{C_0 \left(C_0 
		u^\phi + \Psi u^t\right)}{4\pi g_{RR} 
		g_{\Theta\Theta} u^R} \;, \hspace{1cm}
        T_{{\rm EM},t}^{~~~~~R} = - \Omega_\Psi
                T_{{\rm EM},\phi}^{~~~~~R} \;,
        \label{ttrap}
\end{eqnarray}
where
\begin{eqnarray}
	\Omega_\Psi \equiv - \frac{\Psi(\Theta)}{C_0(\Theta)}
		= \Omega_\Psi(\Theta) \;.
\end{eqnarray}
The corresponding equations for mass conservation, angular momentum 
conservation, and energy conservation are respectively
\begin{eqnarray}
	\frac{\partial}{\partial R}\left(\sqrt{-g}\, \rho_{\rm m} 
		u^R\right) &=& 0 \;;
	\label{massap} \\
	\frac{\partial}{\partial R}\left[\sqrt{-g} \left(\rho_{\rm m} 
		u_\phi u^R + T_{{\rm EM},\phi}^{~~~~~R}\right)
		\right] &=& 0 \;;
	\label{angap} \\
	\frac{\partial}{\partial R}\left[\sqrt{-g} \left(\rho_{\rm m} 
		u_t u^R + T_{{\rm EM},t}^{~~~~~R}\right)
		\right] &=& 0 \;.
	\label{enerap}
\end{eqnarray}

Apparently, when $\Psi = 0$ (i.e., $\Omega_\Psi = 0$), we have $T_{{\rm 
EM},t}^{~~~~~R} = 0$. Then, from Eqs.~(\ref{massap}) and (\ref{enerap})
we have
\begin{eqnarray}
	\frac{\partial u_t}{\partial R} = 0 \;,
\end{eqnarray}
i.e., the specific energy $E = - u_t$ is constant on each stream
foliation surface. Therefore, solutions with constant specific energy 
exist for the general model considered here.

Assuming that $u_t \approx -u^t \approx -1$, $|u^R/(R u^\phi)|\ll 1$,
and $u_\phi\rightarrow\infty$ as $R\rightarrow\infty$, then we have
\begin{eqnarray}
	\left|\sqrt{-g}\, B^R\right|_{R\rightarrow\infty} = 
		\left|\Psi u_\phi\right|_{R\rightarrow\infty} 
		\rightarrow\infty \;, \hspace{1cm}
	\left.\frac{B^2}{4\pi\rho_{\rm m}}\right|_{R\rightarrow\infty} 
		\propto\left.\frac{R^2 \Psi^2}{\sqrt{-g} |u^R|}
		\right|_{R\rightarrow\infty}\rightarrow\infty \;,
\end{eqnarray}
if $\Psi\neq 0$ and $(R^2/\sqrt{-g})_{R\rightarrow\infty}$ is finite
or infinitely large. (For references, $\sqrt{-g} \propto R^2$ for spherical 
coordinates, $\propto R$ for cylindrical coordinates.) To prevent this
to happen, as for the model in the text we must have $\Psi = 0$.

\section{Stationary and Axisymmetric Newtonian Accretion Disks 
Driven by Magnetic Fields}
\label{appb}

The dynamics of a Newtonian, perfectly conducting, and magnetized accretion 
disk is governed by the following equations (\cite{bal98}; here we neglect 
viscosity and resistivity)
\begin{eqnarray}
     \frac{\partial\rho_{\rm m}}{\partial t} + \nabla\cdot(\rho_{\rm m}{\bf v}) 
	     = 0 \;,\hspace{3.4cm}\label{cont} \\
	\rho_{\rm m}\left(\frac{\partial}{\partial t}+ {\bf v}\cdot\nabla
	     \right){\bf v} = -\rho_{\rm m}\nabla\Phi - \nabla\left(p +
		\frac{B^2}{8\pi}\right) + \frac{1}{4\pi} {\bf B}\cdot
		\nabla{\bf B} \;, \label{mom} \\
	\frac{\partial{\bf B}}{\partial t} = \nabla\times({\bf v}\times
	     {\bf B}) \;, \hspace{3.5cm}\label{free}\\
	\nabla\cdot{\bf B} = 0 \;, \hspace{4.4cm}\label{div} 
\end{eqnarray}
where $\rho_{\rm m}$ is the mass density, ${\bf v}$ is the velocity, $\Phi$ is 
the gravitational potential of the central compact object, $p$ is the gas 
pressure, and ${\bf B}$ is the magnetic field. Equation~(\ref{cont})--(\ref{div})
are respectively the continuity, momentum, induction, and divergence-free
equations. 

Let us consider a stationary and axisymmetric Newtonian accretion disk. Similar 
to the relativistic case studied in the text, we adopt spherical coordinates and
assume that in a small neighborhood of the equatorial plane the velocity and 
magnetic field are given by
\begin{eqnarray}
     {\bf v} = v_r {\bf e}_r + v_\phi {\bf e}_\phi \;, \hspace{1cm}
	{\bf B} = B_r {\bf e}_r + B_\phi {\bf e}_\phi \;,
	\label{velo}
\end{eqnarray}
where ${\bf e}_i$ ($i = r, \theta, \phi$) are the unit coordinate vectors, $v_i$ 
and $B_i$ depend only on $r$. Then, Eqs.~(\ref{cont})--(\ref{div}) become 
first-order ordinary differential equations with radius $r$ as the only variable. 
They can be easily solved.

The solutions to Eqs.~(\ref{cont}), (\ref{free}), and (\ref{div}) are
\begin{eqnarray}
     r^2 \rho_{\rm m} v_r = \Psi_{\rm m}/4\pi \;, 
	     \hspace{0.25cm}\label{rrhov} \\
	r\left(v_r B_\phi- v_\phi B_r\right) = C_1 \;, \label{sol1} \\
	r^2 B_r = \Psi_{\rm M}/4\pi \;, \hspace{0.5cm}\label{rbr}
\end{eqnarray}
respectively, where $\Psi_{\rm m}$, $C_1$, and $\Psi_{\rm M}$ are integral
constants. 

Assume that $v_\phi$ remains finite as $r\rightarrow\infty$. Then, from
Eqs.~(\ref{sol1}) and (\ref{rbr}) we have $B_\phi(r\rightarrow\infty) = C_1 /
r v_r$ if $C_1 \neq 0$. If $v_r(r\rightarrow\infty)$ is also finite, then we
find that $\left(B^2/4\pi\rho_{\rm m}\right)_{r\rightarrow\infty} = C_1^2/
\Psi_{\rm m} v_r$,
where $B^2 = B_r^2 + B_\phi^2$. For the case of accretion disks, it is 
reasonable to require that $v_r = 0$ and $B^2/4\pi\rho_{\rm m}$ is finite as 
$r\rightarrow\infty$, therefore we must have $C_1 = 0$, i.e.,
\begin{eqnarray}
     \frac{B_r}{B_\phi} = \frac{v_r}{v_\phi} \label{bvrf} 
\end{eqnarray}
by Eq.~(\ref{sol1}). Equation~(\ref{bvrf}) means that the magnetic field is 
parallel to the velocity field everywhere. 

Substitute Eq.~(\ref{velo}) into Eq.~(\ref{mom}), we obtain
\begin{eqnarray}
     \frac{v_\phi^2}{r} - \frac{1}{2}\frac{dv_r^2}{dr} = 
          \frac{d\Phi}{dr} +\frac{1}{\rho_{\rm m}}\frac{dp}
		{dr} +\frac{1}{8\pi\rho_{\rm m} r^2}\frac{d}{dr}\left(r B_\phi
		\right)^2 \;, \label{mom_r}\\
	v_r\frac{d}{dr}\left(r v_\phi\right) = \frac{B_r}{4\pi\rho_{\rm m}}
	     \frac{d}{dr}\left(r B_\phi\right) \hspace{2.cm} 
		\label{mom_f} 
\end{eqnarray}
in the $r$ and $\phi$-directions near the equatorial plane, respectively. 

From Eqs.~(\ref{rrhov}), (\ref{rbr}), and (\ref{mom_f}), we obtain
\begin{eqnarray}
     \rho_{\rm m} v_r v_\phi = \frac{1}{4\pi} B_r B_\phi + \frac{\Psi_{\rm L}}
		{4\pi r^3} \;, 
	\label{vrbr}
\end{eqnarray}
where $\Psi_{\rm L}$ is the constant of angular momentum flux in the 
radial direction. Substitute Eq.~(\ref{bvrf}) into Eq.~(\ref{vrbr}), we obtain
\begin{eqnarray}
     B_r^2 = 4\pi\rho_{\rm m} v_r^2 \left(1 - \frac{\Psi_{\rm L}}
	     {\Psi_{\rm m}}\frac{1}{L}\right) \;, \hspace{1cm}
     B_\phi^2 = 4\pi\rho_{\rm m} v_\phi^2 \left(1 - \frac{\Psi_{\rm L}}
	     {\Psi_{\rm m}}\frac{1}{L}\right)\;, \label{vfbf}
\end{eqnarray}
where $L \equiv r v_\phi$ is the specific angular momentum of fluid particles.
As $r\rightarrow \infty$ we have $B^2 = B_r^2+ B_\phi^2 = 4\pi\rho_{\rm m} v^2$, 
if $L(r\rightarrow \infty)$ is unbounded. That is, at large radii the magnetic 
energy density is equipartitioned with the kinetic energy density. Note that 
this conclusion crucially depends on the validity of Eq.~(\ref{bvrf}).

From Eqs.~(\ref{bvrf}) and (\ref{mom_f}) we have
\begin{eqnarray}
     \frac{d}{dr}\left(r v_\phi\right)^2 = \frac{1}{4\pi\rho_{\rm m}}
	     \frac{d}{dr} \left(r B_\phi\right)^2 \;.
	\label{deq1}
\end{eqnarray}
Substitute Eq.~(\ref{deq1}) into Eq.~(\ref{mom_r}), we obtain
\begin{eqnarray}
     \frac{d}{dr}\left(\frac{1}{2} v^2 + \Phi\right) = 
	     -\frac{1}{\rho_{\rm m}}\frac{dp}{dr} \;, \label{mom_r2}
\end{eqnarray}
where $v^2 = v_r^2 + v_\phi^2$\,. For a barotropic equation of state $p = 
p(\rho_{\rm m})$, the solution is
\begin{eqnarray}
     \frac{1}{2} v^2 + \Phi + h = \frac{\Psi_{\rm E}}{\Psi_{\rm m}} \;,
	\label{mom3}
\end{eqnarray}
where $\Psi_{\rm E}$ is the constant of energy flux, $h(\rho_{\rm m}) \equiv
\int_0^{\rho_{\rm m}} \rho_{\rm m}^{-1} dp$ is the specific enthalpy.

When $p = 0$, we have $h = 0$, Eq.~(\ref{mom3}) then implies that the specific 
energy of disk particles, $(1/2) v^2 + \Phi$, is constant. This corresponds to 
the solutions of constant specific energy found in the text. 

From Eqs.~(\ref{rrhov}), (\ref{rbr}), (\ref{bvrf}), (\ref{vrbr}), and 
(\ref{mom3}) we can derive the radial momentum equation for the case of 
$p = 0$
\begin{eqnarray}
     v_r^2 = 2\frac{\Psi_{\rm E}}{\Psi_{\rm m}} -2 \Phi - \frac{1}{r^2}
	     \left[\frac{\Psi_{\rm L}/\Psi_{\rm m}}{1 - \left(\Psi_{\rm M}^2/
		16\pi^2 \Psi_{\rm m}\right)\,\left(r^2 v_r\right)^{-1}}\right]^2 \;.
	\label{mom_4}
\end{eqnarray}
Eq.~(\ref{mom_4}) corresponds to the Newtonian limit of the general 
relativistic radial momentum equation~(\ref{floweq}) (with $\Psi = 0$). 
(Note that $-\Psi_{\rm L}/\Psi_{\rm m}$ corresponds to $f_{\rm L}$, 
$-\Psi_{\rm M}^2/16\pi^2 \Psi_{\rm m}$ corresponds to $c_0^2$, $-\Psi_{\rm 
E}/\Psi_{\rm m}$ corresponds to $f_{\rm E} +1$ and satisfies $\left|\Psi_{\rm 
E}/\Psi_{\rm m}\right|\ll 1$.)
A differential form of the radial momentum equation can be derived from the 
differentiation of Eq.~(\ref{mom_4}) with respect to $v_r$ and $r$, as we 
did for the general relativistic case. 

The asymptotic solution for $v_r$ at infinity can be obtained from 
Eq.~(\ref{mom_4}). Suppose $v_r(r\rightarrow\infty) = 0$, then Eq.~(\ref{mom_4}) 
leads to 
\begin{eqnarray}
     \frac{1}{r^2}\left[\frac{\Psi_{\rm L}/\Psi_{\rm m}}{1 - \left(\Psi_{\rm 
		M}^2/16\pi^2 \Psi_{\rm m}\right)\,\left(r^2 v_r\right)^{-1}}\right]^2
	     = 2\frac{\Psi_{\rm E}}{\Psi_{\rm m}} 
	\nonumber
\end{eqnarray}
as $r\rightarrow\infty$. Since the left-hand side cannot be negative, asymptotic 
solutions at infinity exist only if $\Psi_{\rm E}/\Psi_{\rm m} \ge 0$. Note that
$-\Psi_{\rm E}/\Psi_{\rm m}$ is just the specific gravitational binding energy of 
fluid particles when $p =0$ [see Eq.~(\ref{mom3})]. In this paper we do not 
discuss a disk with negative binding energy (i.e. $\Psi_{\rm E}/\Psi_{\rm m}>0$),
thus we will assume $\Psi_{\rm E}/\Psi_{\rm m} =0$ as we discuss asymptotic 
solutions. Then, from Eq.~(\ref{mom_4}), we have
\begin{eqnarray}
     v_r \approx \frac{\Psi_{\rm M}^2}{16 \pi^2 \Psi_{\rm m}} \frac{1}{r^2} \;,
	\label{avr}
\end{eqnarray}
as $r\rightarrow\infty$. The corresponding asymptotic solution for $v_\phi$ is
\begin{eqnarray}
     v_\phi = \left(-v_r^2 - 2\Phi\right)^{1/2} \approx 
	     \left(\frac{2M}{r}\right)^{1/2}  \;, \label{avf}
\end{eqnarray}
where we have taken $\Phi = -M/r$ for the Newtonian potential of a central
compact object of mass $M$. The asymptotic solution for $\rho_{\rm m}$ can be
obtained from Eqs.~(\ref{rrhov}) and (\ref{avr}):
\begin{eqnarray}
     \rho_{\rm m} \approx 4\pi \left(\frac{\Psi_{\rm m}}{\Psi_{\rm M}}
	     \right)^2 \;, \label{arho}
\end{eqnarray}
which is a constant \footnote{There is another asymptotic solution to 
Eq.~(\ref{mom_4}): $v_r = -(2M/r)^{1/2}$, which leads to $v_\phi \propto
r^{-1}$ and $\rho_{\rm m}\propto r^{-3/2}$. However, these do not correspond 
to the asymptotic solutions for an accretion disk since the latter must satisfy 
$\left|v_r/v_\phi\right|\ll 1$ at large radii.}.

The radial magnetic field is always given by $B_r = \Psi_{\rm M}/4\pi r^2$
[Eq.~(\ref{rbr})]. The asymptotic toroidal magnetic field can be obtained 
from Eqs.~(\ref{vfbf}), (\ref{avf}), and (\ref{arho}): $B_\phi \approx \pm
4\pi \left(\Psi_{\rm m}/\Psi_{\rm M}\right)(2M/r)^{1/2}$. Therefore we have
\begin{eqnarray}
     B^2 \approx B_\phi^2 \approx \left(\frac{4\pi\Psi_{\rm m}}{\Psi_{\rm 
	     M}}\right)^2 \frac{2M}{r}\;, \label{ab2}
\end{eqnarray}
as $r\rightarrow\infty$.

All the above asymptotic solutions agree with those listed in Table~\ref{tab1}
for the general relativistic model. (Note that $\Psi_{\rm m}$ corresponds to
$F_{\rm m}$, and $\Psi_{\rm M}$ corresponds to $4\pi C_0$.) Certainly this has 
to be true since at large distances from the central black hole the dynamics 
becomes Newtonian.


\clearpage
\begin{table}
\vspace{4cm}
\caption{\label{tab1}
Asymptotic solutions corresponding to $\Psi = 0$ and $E = 1$.  
Symbols: $u^r$, radial component of the four-velocity; $\Omega\equiv 
u^\phi/u^t$, angular velocity; $\rho_{\rm m}$, mass-energy density; $B^2 
= B_a B^a$ ($\rho_{\rm m}$ and $B^2$ are measured by an observer 
comoving with the flow); $L = u_\phi$, specific angular momentum of fluid 
particles; $E = -u_t$, specific energy; $\zeta$, ratio of electromagnetic
energy density to kinetic energy density [Eq.~(\ref{zeta})], measured by 
a LNRF observer; $u_{\rm H}^r$ is given by Eq.~(\ref{urh}); $L_{\rm H}$ 
by Eq.~(\ref{lh}); $\zeta_{\rm H}$ by Eq.~(\ref{zetah}) ($\Omega_\Psi= 0$, 
$f_{\rm E} = -1$); $\Omega_{\rm H}$, angular velocity of the black hole 
horizon. Supersonic/subsonic motion refers to the fact that $u_r u^r$ is 
greater/smaller than $c_{\rm A}^2/\left(1-c_{\rm A}^2\right)$;
$u_r u^r = c_{\rm A}^2/\left(1-c_{\rm A}^2\right)$ defines the fast critical 
point [Eq.~(\ref{rf})]. Note that $c_0 = C_0/\sqrt{-F_{\rm m}}\,$.}
\vspace{0.4cm}
\begin{ruledtabular}
\begin{tabular}{lccccccc}
Region & $u^r$ & $\Omega$ & $\rho_{\rm m}$ & $B^2$ & $L$ & $\zeta$
& Motion \\
\hline
$r\rightarrow r_{\rm H}$ & $u_{\rm H}^r$ & $\Omega_{\rm H}$   
	& $\frac{F_{\rm m}}{4\pi r_{\rm H}^2 u_{\rm H}^r}$ & 
	$\frac{2M}{r_{\rm H}}\left(\frac{C_0}{r_{\rm H}^2 u_{\rm H}^r}
	\right)^2$ & $L_{\rm H}$ & $\zeta_{\rm H}$ & supersonic   \\
$r\rightarrow\infty$ & $-\frac{c_0^2}{r^2}$ & $\left(\frac{2M}{r^3}
	\right)^{1/2}$ & $\frac{1}{4\pi} \left(\frac{F_{\rm m}}
	{C_0}\right)^2$ & $\frac{2M}{r}\left(\frac{F_{\rm m}}
	{C_0}\right)^2$ & $(2M r)^{1/2}$ & 
	$1$ & subsonic  \\
\end{tabular}
\end{ruledtabular}
\end{table}

\clearpage

\clearpage

\clearpage

\clearpage

\clearpage

\clearpage

\clearpage

\clearpage

\clearpage

\clearpage

\clearpage

\clearpage

\clearpage

\end{document}